\documentclass[10pt,letterpaper]{article}
\usepackage[top=0.85in,left=2.75in,footskip=0.75in]{geometry}
\usepackage{amsmath,amssymb}
\usepackage{bm}

\usepackage{float}
\usepackage{booktabs}
\usepackage{multicol}

\def\ie{{\frenchspacing\it i.e.}}
\def\eg{{\frenchspacing\it e.g.}}

\def\etc{{\frenchspacing\it etc.}}

\def\relu{{\rm ReLU}}

\def\I{{\bf I}}

\def\D{{\bf D}}
\def\E{{\bf E}}
\def\Ell{{\mathcal L}}

\def\M{{\bar\N}}
\def\Mhat{\M}
\def\N{{\bf N}}
\def\Nbar{{\bar N}}
\def\Nbarhat{{\Nbar}}
\def\Nbarhatb{{{\bar\N}}}

\def\R{{\bf R}}

\def\U{{\bf U}}
\def\V{{\bf V}}
\def\W{{\bf W}}

\def\ith{i^{\rm th}}
\def\jth{j^{\rm th}}
\def\kth{k^{\rm th}}

\def\spose#1{\hbox to 0pt{#1\hss}}
\def\simlt{\mathrel{\spose{\lower 3pt\hbox{$\mathchar"218$}}
     \raise 2.0pt\hbox{$\mathchar"13C$}}}
\def\simgt{\mathrel{\spose{\lower 3pt\hbox{$\mathchar"218$}}
     \raise 2.0pt\hbox{$\mathchar"13E$}}}
\def\simpropto{\mathrel{\spose{\lower 3pt\hbox{$\mathchar"218$}}
     \raise 2.0pt\hbox{$\propto$}}} 
     
\def\beq#1{\begin{equation}\label{#1}}
\def\eeq{\end{equation}}
\def\beqa#1{\begin{eqnarray}\label{#1}}
\def\eeqa{\end{eqnarray}}
\def\eq#1{equation~(\ref{#1})}	

\def\eqnum#1{~(\ref{#1})}

\def\fig#1{FIG.~\ref{#1}}
\def\Fig#1{FIG.~\ref{#1}}

\def\tabl#1{Table~\ref{#1}}
\def\Tabl#1{Table~\ref{#1}}

\usepackage{changepage}

\usepackage[utf8x]{inputenc}

\usepackage{textcomp,marvosym}

\usepackage{cite}

\usepackage{nameref,hyperref}

\usepackage{microtype}
\DisableLigatures[f]{encoding = *, family = * }

\usepackage[table]{xcolor}

\usepackage{array}

\newcolumntype{+}{!{\vrule width 2pt}}

\newlength\savedwidth

\newcommand\thickhline{\noalign{\global\savedwidth\arrayrulewidth\global\arrayrulewidth 2pt}%
\hline
\noalign{\global\arrayrulewidth\savedwidth}}

\raggedright
\usepackage[aboveskip=1pt,labelfont=bf,labelsep=period,justification=raggedright,singlelinecheck=off]{caption}

\makeatletter
\renewcommand{\@biblabel}[1]{\quad#1.}
\makeatother

\usepackage{lastpage,fancyhdr,graphicx}
\usepackage{epstopdf}
\fancyheadoffset[L]{2.25in}
\fancyfootoffset[L]{2.25in}
\begin{document}
\vspace*{0.2in}

\begin{flushleft}
{\Large
\textbf\newline{Machine-Learning media bias} 
}
\newline
\\
Samantha D'Alonzo\textsuperscript{1},
Max Tegmark\textsuperscript{1}
\\
\bigskip
\textbf{1}~Dept.~of Physics and Institute for AI \& Fundamental Interactions,\\
Massachusetts Institute of Technology, Cambridge, MA, USA; tegmark@mit.edu\\
\bigskip
\end{flushleft}
\section*{Abstract}
We present an automated method for measuring media bias.
Inferring which newspaper published a given article, based only on the frequencies with which it uses different phrases, leads to a conditional probability distribution whose analysis lets us automatically map newspapers and phrases into a bias space.
By analyzing roughly a million articles from roughly a hundred newspapers for bias in dozens of news topics, our method maps newspapers into a two-dimensional bias landscape that agrees well with previous
bias classifications based on human judgement. One dimension can be interpreted as traditional left-right bias, the other as establishment bias. This means that although news bias is inherently political, its measurement need not be.
\section*{Author summary}
Many argue that news media they dislike are biased, while their favorite news sources aren't. Can we move beyond such subjectivity and measure media bias objectively, from data alone? 
Our answer is a resounding ``yes" after analyzing roughly a million articles from roughly a hundred newspapers.
By simply aiming to machine-learn which newspapers published which articles, 
hundreds of politically charged phrases such as {\it ``undocumented immigrant"} and {\it ``illegal immigrant"} are auto-identified, whose relative frequencies enable us to map all newspapers into a two-dimensional media bias landscape. These data-driven results agree well with human-judgement classifications of left-right bias and establishment bias.

\section*{Introduction}
\label{IntroSec}

Political polarization has increased in recent years, both in the United States and internationally \cite{wilson_polarization_2020}, 
with pernicious consequences for democracy and its ability to solve pressing problems \cite{mccoy2018polarization}
It is often argued that such polarization is stoked by the media ecosystem, with machine-learning-fueled filter bubbles \cite{pariser2011filter} increasing the demand for and supply of more biased media.
Media bias is defined by \cite{shultziner_distorting_2021}
as favoring, disfavoring, emphasizing or ignoring certain political actors, policies, events, or topics in a way that is deceptive toward the reader,
and can be accomplished through many different techniques.

In response, there has been significant efforts to protect democracy by studying and flagging media bias. However, there is a widespread perception that fact-checkers and bias-checkers can themselves be biased and lack transparency \cite{brandtzaeg2017trust}. It is therefore of great interest to develop objective and transparent measures of bias that
are based on data rather than subjective human judgement calls.
Early work in this area is reviewed in \cite{groseclose_socialscience_2005},
and is mainly qualitative, manual, or both. While this has produced interesting findings on biased coverage of, \eg, protests \cite{mccarthy_assessing_2008} and terrorism \cite{papacharissi_news_2008}, the manual nature of these methods limits their scalability and feasibility for real-time bias monitoring in the digital age.

Advances in machine learning (ML) raise the possibility of bias detection that is transparent and scalable by virtue of being automated, with little or no human intervention. Early efforts in this direction have shown great promise, as reviewed in \cite{hamborg_media_2020}. 
For example, various ML natural language processing (NLP) techniques have been employed to discover bias-inducing words from articles in four German newspapers \cite{spinde_media_2020} and six 20th Century Dutch newspapers \cite{wevers_using_2019}. ML NLP techniques have also been used to detect gender bias in sports interviews \cite{fu_tie-breaker_2016}, to detect political bias in coverage of climate change\cite{schuldt_global_2011}, to identify trolling in social media posts \cite{liu2019finding}, and to analyze bigram/trigram frequencies in the U.S. congressional record\cite{gentzkow_what_2010}.
Although these studies have been successful, they have typically involved relatively small datasets or hand-crafted features, making it timely and interesting to further pursue automated media bias detection with larger datasets and broader scope. This is the goal of the present paper.

\begin{figure}[t]
\caption{ 
{\bf Generalized principal components for articles about BLM.} The colors and sizes of the dots were predetermined by external assessments and thus in no way influenced by our data. The positions of the dots thus suggest that the horizontal axis can be interpreted as the traditional left-right bias axis, here automatically rediscovered by our algorithm
directly from the data.
}
\hglue-5.89cm\includegraphics[scale=.58]{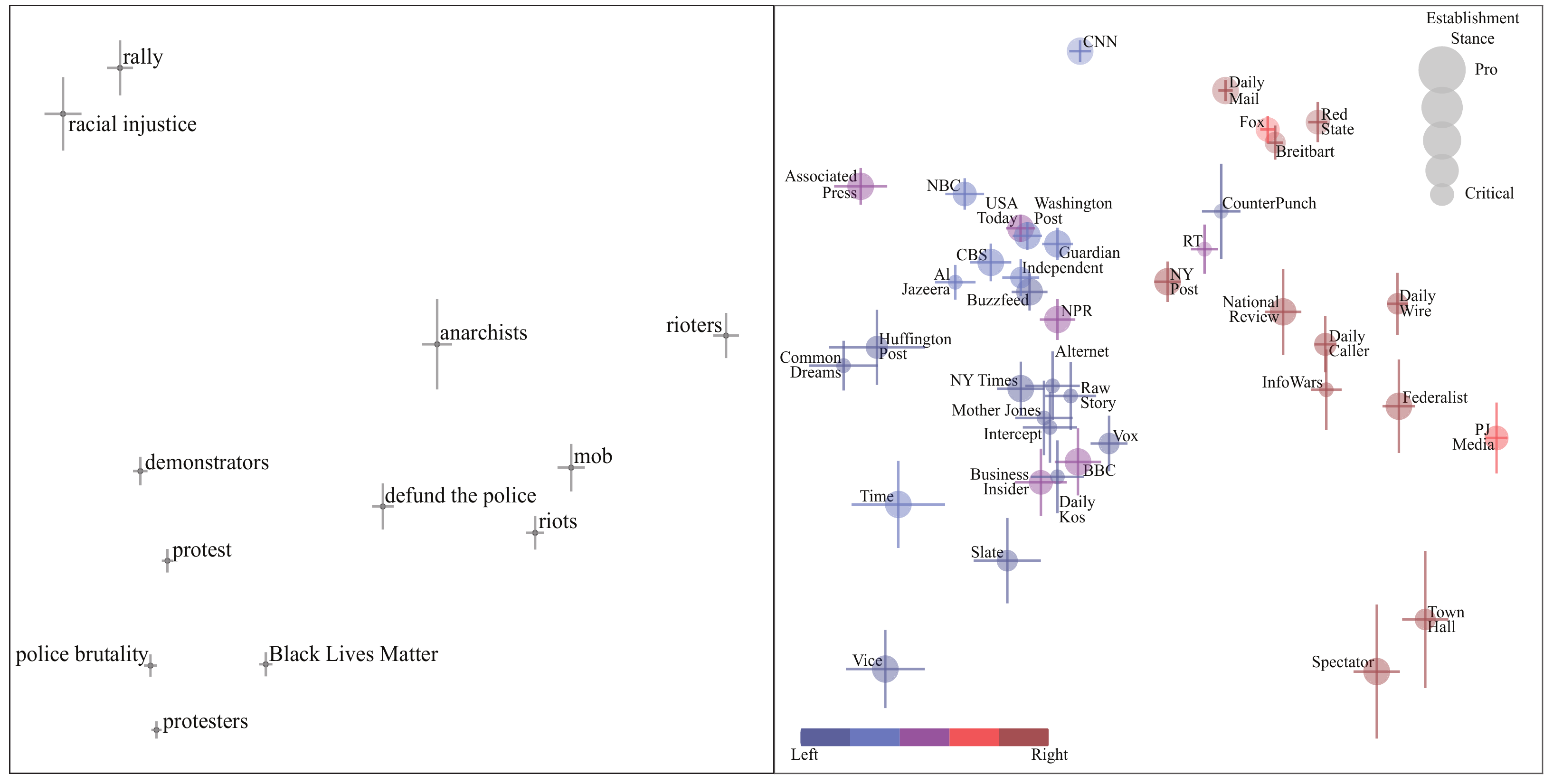}
\label{BLMfig}
\end{figure}

Specifically, we will use a dataset containing roughly a million articles from about 100 different newspapers to study {\it phrase bias} \cite{hamborg_automated_2019}, \ie, the bias allowing a machine-learning algorithm to predict which newspaper published an article 
merely from how often it uses certain phrases.
As illustrated in \fig{BLMfig}, for instance, articles about the Black Lives Matter (BLM) topic 
that refer to ``demonstrators" and ``rioters" are likely to be from media on the political left and right, respectively. Our goal is to make the bias-detection algorithm as automated, transparent and scalable as possible, so that biases of phrases and newspapers are machine-learned rather than input by human experts.  For example, the horizontal positions of phrases and newspapers in \fig{BLMfig}, which can be interpreted in terms of left-right bias, were computed directly from our data, without using any human input as to how various phrases or media sources may be biased.

The rest of this paper is organized as follows. The Methods section describes our algorithm for automatically learning media bias from an article database, including a generalization of principal component analysis tailored for phrase frequency modeling. The Results section shows our findings for the most biased topics, and identifies a two-dimensional bias landscape that emerges from how bias correlates across topics, with left-right stance and establishment stance as its two bias axes.
The Conclusions section summarizes and discusses our findings.

\section*{Methods }

In this section, we present our method for automated bias detection.
We first describe how we automatically map both phrases, meaning monograms, bigrams, or trigrams, and newspapers into a $d$-dimensional bias space using phrase statistics alone, then present our method for phrase selection. 

\subsection*{Generalized SVD-modeling of phrase statistics}

Given a set of articles from $n$ different media sources, we begin by counting occurrences of $m$ phrases (say ``fetus'', ``unborn baby", \etc).
We arrange these counts into an 
$m\times n$ matrix $\N$ of natural numbers $N_{ij}\ge 0$ encoding
how many times the $\ith$ phrase occurs in the $\jth$ media source. 
We model $N_{ij}$ as a random variable drawn from a Poisson distribution 
whose mean $\Nbar _{ij}$ (the average number of times the phrase occurs) is non-negative and depends both on the phrase $i$ and the media source $j$:
\beq{PoissonProbEq}
P(N_{ij}=k)=e^{-\Nbar _{ij}}{\Nbar _{ij}^k\over k!}.
\eeq
Our goal is to accurately model this matrix $\M$ in terms of biases that link phrases and newspapers. Specifically, we wish to approximate either $\M$ (or, alternatively, its logarithm) as a low-rank matrix $\Mhat$, as in Singular-Value Decomposition (SVD) \cite{eckart_approximation_1936}:
$$\Nbar_{ij}\approx\Nbarhat_{ij} \equiv \sum_{k=1}^r w_k U_{ik}V_{jk},$$
where the rank $r<\min(m,n)$.
Without loss of generality, we can choose $\U$ and $\V$ to be orthogonal matrices ($\U\U^t=\I$, $\V\V^t=\I$) and $w_k>0$.

Singular-value decomposition (SVD) corresponds to minimizing the mean-squared-error 
loss function
$L_{\rm SVD}=||\Mhat-\M||_2^2$. Although SVD is easy to compute and interpret mathematically, it is poorly matched to our media bias modeling problem for two reasons. 
 First of all, it will in some cases predict negative phrase counts $\Nbarhat_{ij}$, which of course makes no sense as a language model.
Second, it implicitly gives equal weight to fitting every single number $\Nbar _{ij}$, even though some are measured much more accurately than others from the data (the Poission error bar is  
 $\sqrt{\Nbar _{ij}}$ and phrase counts can differ from one another by orders of magnitude). 
To avoid these shortcomings, we choose to not minimize the SVD loss, but to instead maximize the Poisson likelihood
\beq{PoissonLeq}
L_{\rm Poisson}=\prod_{ij} e^{-\Nbar _{ij}}{\Nbar _{ij}^{N _{ij}}\over N _{ij}!},
\eeq
\ie, the likelihood that our model produces the observed phrase counts $\N$. Numerically, it is more convenient to maximize its logarithm 
\beq{PoissonlnLeq}
\Ell\equiv \ln L_{\rm Poisson}= \sum_{ij} N _{ij}\ln({\Nbar _{ij}) - \ln(N _{ij}!})- \Nbar _{ij}
\approx
-\sum_{ij}\left[\Nbar_{ij}  +  N _{ij} \ln\left({N _{ij}\over e\Nbar_{ij}}\right)\right].
\eeq
The approximation in the last step uses
 Stirling's approximation $\ln(k!)\approx k\ln(k/e)$, and we use it for numerical speedup
 only when $\Nbar _{ij}>50$.
 To avoid the aforementioned problems with forbidden negative $\Nbar$-values, we 
 try two separate fits and select the one that fits the data better (gives a higher Poisson likelihood):
\beqa{NonNegativeEq}
\Nbarhat_{ij}&\equiv&\relu\left[\sum_{k=1}^r w_k U_{ik}V_{jk}\right],\\
\Nbarhat_{ij}&\equiv&\exp\left[ \sum_{k=1}^r w_k U_{ik}V_{jk}\right],\label{expNeq}
\eeqa
where $\relu(x)=x$ if $x\ge 0$, vaninishing otherwise.
In our numerical calculations in the Results section, we find that the second fit performs better most of the time, but not always.

We determine the best fit by selecting the desired rank $r$ (typically $r=3$) and numerically minimizing the loss function
$\Ell\equiv -\ln L_{\rm Poisson}$ over the fitting parameters $w_k$, $U_{ik}$ and $V_{jk}$.
We do this using the gradient-descent method method implemented in {\it scipy.optimize}
\cite{virtanen2020scipy},
which is greatly accelerated by the following exact formulas for $\nabla\Ell$ that follow from equations\eqnum{PoissonlnLeq}~and\eqnum{NonNegativeEq}: 
\beqa{GradientEq}
\nabla_\U \Ell&=&\D\V\W,\\
\nabla_{\V^t} \Ell&=&\W\U^t\D,\\
{\partial\Ell\over\partial w_i}&=&(\U^t\D\V)_{ii},
\eeqa
where 
\beq{DdefEq}
D_{ij} = \left(1-{N_{\ij}\over\Nbarhat_{ij}}\right)\theta(\Nbarhat_{ij}),
\eeq
$\W$ is the diagonal matrix with $W_{kk}=w_k$, 
and $\theta$ is the Heaviside step function defined by $\theta(x)=1$ if $x>0$, vanishing otherwise.
For the exponential parametrization of \eq{expNeq}, these formulas are identical except that $\D=\Nbarhatb-\N$.
Once the numerical optimization has converged and determined $\Mhat$, we use the aforementioned freedom to ensure that $\U$ and $\V$ are orthogonal matrices and $w_k\ge 0$.

\label{method}
\subsection*{Data} \label{data}

\begin{figure}[tb]
\caption{{\bf Media legend for generalized principle component plots}}
\hglue-5.7cm\includegraphics[scale = .8]{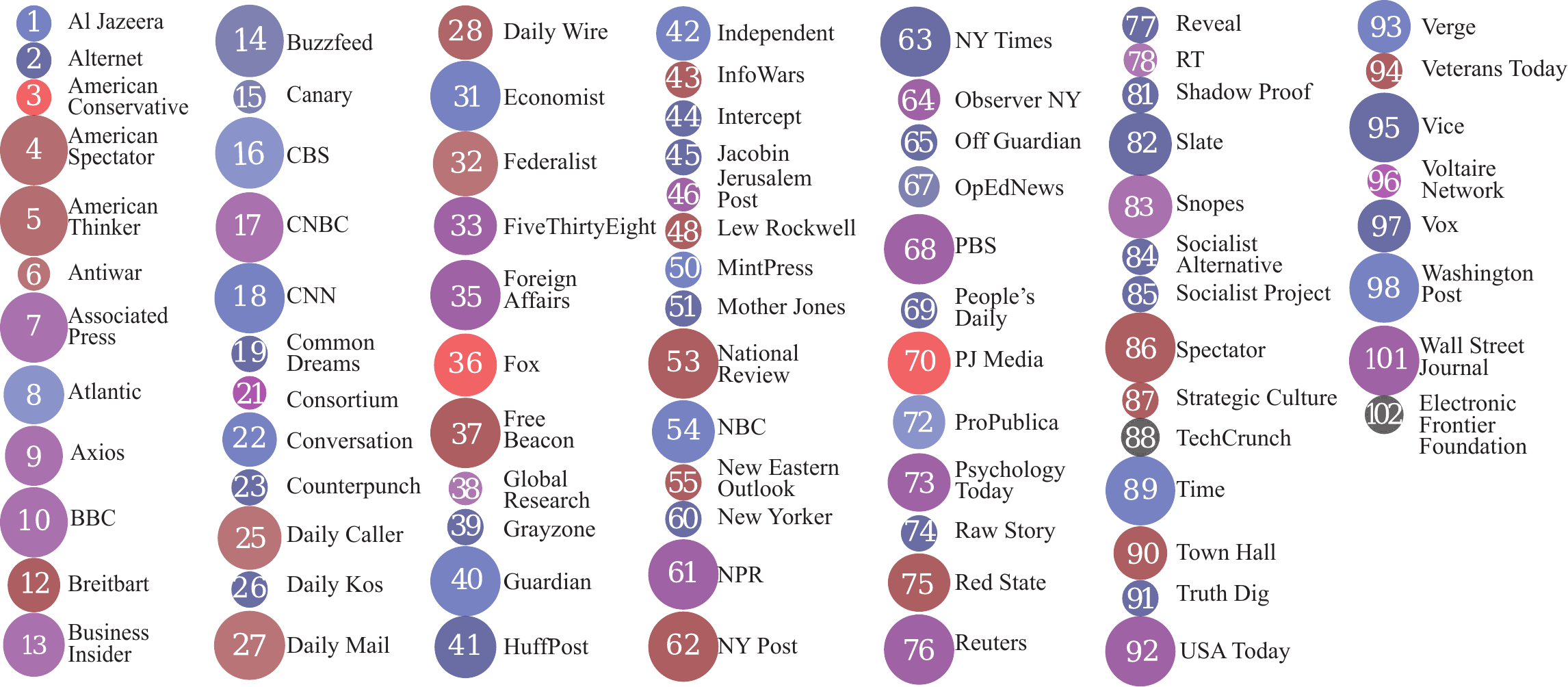}
\label{legend}
\end{figure}

Using the open-source {\it Newspaper3k} software \cite{ou-yang_newspaper3k_nodate}, we scraped and downloaded a total of 3,078,624 articles published between January 2019 and December 2020 from 100 media sources chosen to include the largest US newspapers as well as a broad diversity of political stances.
The 83 newspapers appearing in our generalized SVD bias figures below are listed in in \fig{legend} and the correlation analysis at the end also includes articles from {\it Defense One} and {\it Science}.

The downloaded article text was minimally pre-processed before analysis. 
All text in ``direct quotes" was removed from the articles, since we are interested in biased phrases use by journalists, not by their quoted sources. We replaced British spelling of common words (\eg, favourite, flavour) with  American spelling (favorite, flavor) to erase spelling-based clues as to which newspaper an article is from.
Non-ASCII characters were replaced by their closest ASCII equivalent.  
Text was stripped of all punctuation marks except periods, which were removed only when they did not indicate end-of-sentence --- for example,  ``M.I.T." would become ``MIT". End-of-sentence periods were replaced by ``PERIOD" to avoid creating false bigrams and trigrams containing words not in the same sentence.
Numerals were removed unless they were ordinals ($1^{st}$, $17^{th}$), in which case they were replaced with equivalent text (first, seventeenth). The first letter of each sentence was lower-cased, but all other capitalization was retained. 
 We discarded any articles containing fewer than ten words after the aforementioned preprocessing.

\subsection*{Extraction of discriminative phrases}

We auto-classified the articles by topic using the open-source {\it MITNewsClassify} package from \cite{wongprommoon_mit-news-classify_nodate}. For each of the topics mentioned below (covered in 779,174 articles), we extracted discriminative phrases by
 first extracting the 10,000 most common phrases, then ranking, purging and merging this phrase list as described below.

\subsubsection*{Automatic purge}

To avoid duplication, we deleted subsumed monograms and bigrams from our phrase list:
we deleted all monograms that appeared in a particular bigram more than 70\% of the time and all bigrams that appeared in a particular trigram more than 70\% of the time. For the BLM topic, for example, ``tear" was deleted because if appeared in ``tear gas" 87\% of the time. 

Next, all phrases were sorted in order of decreasing information score
\beq{InfoScoreEq}
I_i\equiv\sum_j P_{ij}\log_2 {P_{ij}\over P_{i\cdot} P_{\cdot j}},
\eeq
where $P_{ij}\equiv N_{ij}/N_{\cdot\cdot}$ is the aforementioned $\N$-matrix rescaled as a joint probability distribution over phrases $i$ and newspapers $j$, and replacing an index by a dot denotes that the index is summed over; for example, 
$N_{\cdot\cdot}$ is the total number of phrases in all the articles considered.
The mutual information between phrases and articles is  $\sum_i I_i$, which can be interpreted as how many bits of information 
we learn about which newspaper an article is from by looking at one of its phrases.
The information scores $I_i$ can thus be interpreted as how much of this information the $\ith$ phrase contributes. Phrases are more informative both if they are more common and if their use frequency varies more between newspapers.

We remove all phrases where more than 90\% of all occurrences of the phrase are from a single newspaper.
These ``too good" phrases commonly reference journalist names or other things unique to newspapers but not indicative of political bias. For example, CNBC typically labels its morning news and talk program \textit{Squawk Box}, making the phrase \textit{Squawk Box} useful for predicting that an article is from CNBC but not useful for learning about media bias. 
To further mitigate this problem, we created a black list of newspaper names, journalist names, other phrases uniquely attributable to a single newspaper, and generic phrases that had little stand-alone meaning in our context (such as ``article republished"). Phrases from this list were discarded for all topics.
Phrases that contained \textit{PERIOD} were also removed from consideration. 
Just as we discarded direct quotes above, we also removed all phrases that contained ``said" or ``told"  because they generally involved an indirect quote. 
Once this automatic purge was complete, the 1,000 remaining candidate phrases with the highest information scores were selected for manual screening as  described in the next section. 

\subsubsection*{Manual purge and merge}

To be included in our bias analysis, phrases must meet the following criteria:
\begin{enumerate}
\item {\bf Relevance:}
\begin{itemize}
\item In order to be relevant to a topic, a phrase
must not be a very common one that has ambiguous stand-alone meaning.  For example, the phrase ``social media" could be promoting social media pages, as in ``Follow us on social media", or referencing a social media site. For simplicity, such common phrases with multiple meanings were excluded. Note that longer phrases (bigrams or trigrams) that contained such shorter phrases (monograms or bigrams) could still be included, such as ``social media giants" in the \textit{tech censorship topic}.
\item A phrase is allowed to occur in multiple topics (for example, ``socialism" is relevant to both the \textit{Venezuela} and \textit{Cuba} topics), but a sub-topic is not.
For example, phrases related to the sub-topic {\it tech censorship in China} were excluded from both the \textit{tech censorship} and \textit{China} topics because they were relevant to both.  
\end{itemize}
\item {\bf Uniqueness:}
Since there was minimal pre-processing, many phrases appear with different capitalizations or conjugations. In some cases, only one of the phrase variations was included and the others were discarded. In other cases, all variations were included because they represented a meaningful difference. These choices were made on a case by case basis, with a few general rules. 

If both a singular and plural version of a word were present, only the more frequent variant was kept.
If phrases were differentially capitalized (for example ``big tech" and ``Big Tech"), we kept both if they landed more than two standard deviations apart in the generalized principal component plot, otherwise we kept only the most frequent variant.
If phrases were a continuation of one another, such as ``Mayor Bill de" and ``Bill de Blasio", the more general phrase was included. In this case, ``Bill de Blasio" would be included because it does not contain an identifier. If there was no identifier, the more informative phrase was kept: for example, discarding ``the Green New" while keeping ``Green New Deal".
\item {\bf Specificity:} Phrases must be specific enough to stand alone. A phrase was deemed specific if the phrase could be interpreted without context or be overwhelmingly likely to pertain to the relevant topic. This rules out phrases with only filler words (\eg, ``would like", ``must have") and phrases that are too general (e.g. ``politics").
\item { \bf Organize Subtopics (if needed): } Some topics were far larger and broader than others. For example, \textit{finance} contained many natural subtopics, including \textit{private finance} and \textit{public finance}. If natural subtopics appeared during the above process, the parent topic was split into subtopics. If topics were small and specific, such as \textit{guns}, no such additional manual processing was performed. 
\item {\bf Edge cases:} There were about a dozen cases on the edge of exclusion based on the above criteria, for which the include/exclude decision was based on a closer look at both the underlying data and the phrase error bar emerging from the principal component analysis. Most of these phrases were excluded for occurring only in a single newspaper for stylistic reasons. When necessary, we examined the use of the phrase in context by reading a random sample of 10 articles in our database containing the phrase. 
\end{enumerate}

\section*{Results}

In this section, we present the results of applying our method to the aforementioned 779,174-article dataset.
We will first explore how the well-known left-right media bias axis can be auto-discovered. We then identify a second bias axis related to establishment stance, and conclude this section by investigating how bias correlates across topics.

\subsection*{Left-Right Media Bias}

\begin{table}[t]
\begin{adjustwidth}{-2.25in}{0in} 
\centering
\caption{BLM phrase bias: the average number of occurrences per article of certain phrases is seen to vary strongly between media sources.}
\begin{tabular}{|l|r|r|r|r|r|r|} \hline
\textbf{Phrase} & \textbf{PJ Media} & \textbf{Breitbart} & \textbf{Fox News} & \textbf{Washington Post} & \textbf{NY Times} & \textbf{Counterpunch}\\ 
 \thickhline
{riots} & $.97 \pm .13$ & $.19 \pm .03$ & $.12 \pm .02$ & $.07 \pm .02$ & $.02 \pm .01$ & $.37 \pm .17$ \\ \hline
{mob} & $.43 \pm .09$ & $.10 \pm .02$ & $.04 \pm .01$ & $.01 \pm .01$ & $.01 \pm .01$ & $.22 \pm .13$ \\ \hline
{anarchists} & $.06 \pm .03$ & $.17 \pm .03$ & $.03\pm .01$ & $.01 \pm .01$ & $.01 \pm .01$ & $1.33 \pm .32$ \\ \hline
{protests}& $.70 \pm .11$ & $1.02 \pm .07$ & $.55 \pm .03$ & $.86 \pm .06$ & $.35 \pm .04$ & $1.28 \pm .31$ \\ \hline
{demonstrators}& $.22 \pm .06$ & $.28 \pm .04$ & $.15 \pm .02$ & $.23 \pm .03$ & $.08 \pm .02$ & $.26 \pm .14$ \\ \hline
{rally}& $.06 \pm .03$ & $.07  \pm .04$ & $.04 \pm .02$ & $.08 \pm .02$ & $.04 \pm .01$ & $.16 \pm .11$ \\ \hline
{defund the police}& $.22 \pm .06$ & $.19 \pm .02$ & $.08 \pm .01$ & $.04 \pm .01$ & $.02 \pm .01$ & $.10 \pm .09$ \\ \hline
\end{tabular}
\label{BLM_counts}
\end{adjustwidth}
\end{table}

We begin by investigating the Black Lives Matter (BLM) topic, because it is so timely. The BLM Movement swept across the USA in the summer of 2020, prompting media coverage from newspapers of varied size and political stance. We first compute the aforementioned $\N$-matrix describing phrase statistics; $N_{ij}$ is how many times the $i^{\rm th}$ phrase was mentioned in the $j^{\rm th}$ newspaper. We have made this and all the other $\N$-matrices computed in this paper are available online\footnote{Our $\N$-matrices, phrase lists {\etc} are available at \url{https://space.mit.edu/home/tegmark/phrasebias.html}.}. Table~\ref{BLM_counts} shows a sample, 
rescaled to show the number of occurrences per article,
revealing that the frequency of certain phrases varies dramatically between media sources. 
For example, we see that ``riots" is used about 60 times more frequently in PJ Media than in the NY Times, which prefers using ``protests". 

As described in the previous section, our generalized principal component analysis 
attempts to model this $\N$-matrix in terms of biases that link phrases and newspapers.
The first component (which we refer to as component 0) tends to model the obvious fact that some phrases are more popular in general and some newspapers publish more articles than others, so we plot only the next two components (which we refer to as 1 and 2) below.
BLM components 1 and 2 are shown in \fig{BLMfig}, corresponding to the horizontal and vertical axes: the phrase panel (left) plots $U_{i1}$ against $U_{i2}$ for each phrase $i$
and the media panel (right) plots $V_{j1}$ against $V_{j2}$ for each media source $j$. The bars represent 1 standard deviation error bars computed using the Fisher information matrix method. To avoid clutter, we only show phrases occurring at least 200 times and newspapers with at least 200 occurrences of our discriminative phrases; for topics with fewer than 15,000 articles, we drop the phrase threshold from 200 to 100.

In the media panel, the dots representing newspapers are colored based on external left-right ratings and scaled based on external pro-critical establishment ratings (which crudely correlates with newspaper size)\footnote{The colors of the media dots reflect the left-right classification of media from \cite{noauthor_allsides_nodate} into the five classes ``left", ``lean left", ``center", ``lean right" and ``right". The  sizes of the media dots reflects the establishment stance classification
from \cite{noauthor_improve_nodate} which is based on the Swiss Policy Research Media Navigator classification \cite{noauthor_media_nodate} and Wikipedia's lists 
of left, libertarian and right alternative media, attempting to quantify the extent to which
a news source normally accepts or challenges claims by powerful entities such as the government and large corporations.}.
It is important to note that the colors and sizes of the dots were predetermined by external assessments and thus in no way influenced by the $\N$-matrices that form the basis of our analysis in this paper. 
It is therefore remarkable that \fig{BLMfig} reveals a clear horizontal color separation,
suggesting that the first BLM component (corresponding to the horizontal axis) 
can be interpreted as the well-known left-right political spectrum.

\begin{figure}[tb]
\caption{{\bf Valent synonyms reflecting left-right bias:}
Each row shows phrases that can be used rather interchangeably, 
with a horizontal position reflecting
where our automated algorithm placed them on the left-right bias axis.}
\hglue-1.6cm 
\includegraphics[scale = .9]{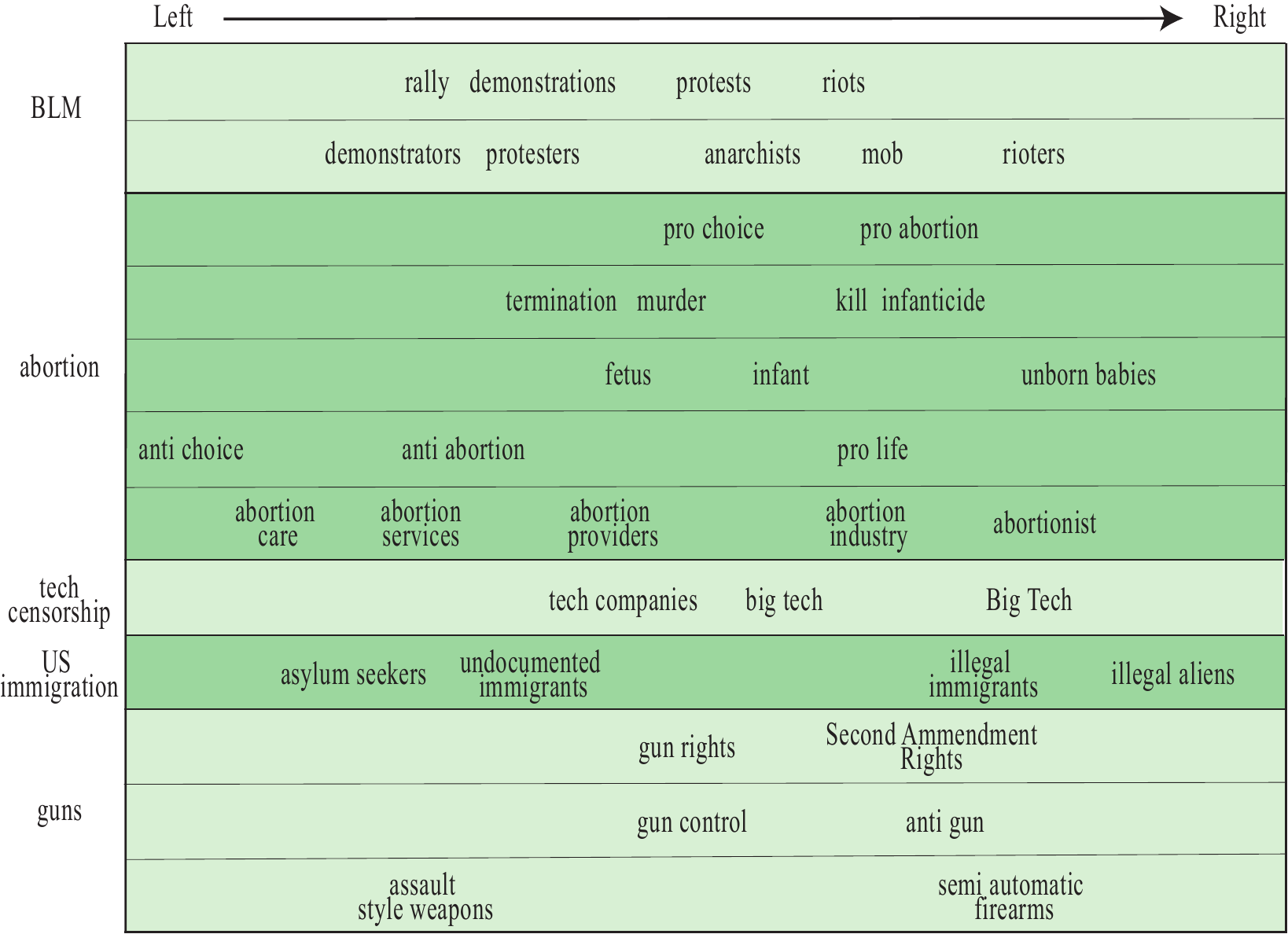}
\label{phrase_plots}
\end{figure}

\subsubsection*{Phrase bias and valent synonyms}

As described in the {\it Methods} section, the phrases appearing in  \Fig{BLMfig} (left panel) were selected by our algorithm as the ones that best discriminated between different newspapers. We see that they typically carry implicit positive or negative valence. 
Looking at how these phrases are used in context reveals that some of them form groups of phrases that can be used
rather interchangeably, \eg, ``protests" and ``riots".
For example, a June 8 2020 New York Times article reads 
{\it ``Floyd's death triggered major \textbf{protests} in Minneapolis and sparked rage across the country"} \cite{nyt_george} while a June 10 2020 Fox News article mentions {\it ``The death of George Floyd in police custody last month and a series of \textbf{riots} that followed in cities across the nation" } \cite{noauthor_george_nodate}.
The $x$-axis in \Fig{BLMfig} is seen to automatically separate this pair, with ``protests" on the left and ``riots" on the right, with newspapers (say NY Times and PJ Media) similarly being left-right separated in the right panel according to their relative preference for these two phrases. 
\Fig{phrase_plots} shows many such groups of emotionally loaded near-synonyms for both BLM and other topics.
In many cases, we see that such a phrase group can be viewed as falling on a linguistic valence spectrum from positive  (euphemism) to neutral (orthophemism) to negative (dysphemism).

\subsubsection*{The nutpicking challenge}

\Fig{BLMfig} is seen to reveal a clean, statistically significant split between almost all left-leaning and right-leaning newspapers. The one noticeable exception is {\it Counterpunch}, whose horizontal placement shows it breaking from its left-leaning peers on BLM coverage. 
A closer look at the phrase observations reveals that this interpretation is misleading, and an artifact of some newspapers placing the same phrase in contexts where it has opposite valence. 
For example, a {\it Counterpunch} article treats the phrase ``defund the police" as having positive valence by writing 
{\it ``the advocates of \textbf{defund the police} aren't fools. They understand that the police will be with us but that their role and their functions need to be dramatically rethought"} \cite{noauthor_police_2020}.
In contrast, right-leaning PJ Media  treats ``defund the police" as having negative valence in this example:
{\it ``If you're a liberal, whats not to like about the slogan \textbf{defund the police}? It's meaningless, it's stupid, it's dangerous, and it makes you feel good if you mindlessly repeat it"} \cite{moran_bloody_nodate}. 
This tactic is known as {\it nutpicking}: picking out and showcasing what your readership perceives as the nuttiest statements of an opposition group as representative of that group. 

In other words, whereas most discriminative phrases discovered by our algorithm have a context-independent valence (``infanticide" always being negative, say), some phrases are bi-valent in the sense that their valence depends on how they are used and by whom.
We will encounter this challenge in many of the news topics that we analyze; for example, most U.S.~newspapers treat ``socialism" as having negative valence, and as a result, the arguably most socialist-leaning newspaper in our study, {\it Socialist Alternative}, gets mis-classified as right-leaning because of its frequent use of ``socialism" with positive connotations. For example, for the Venezuela topic, 
{\it Socialist Project} uses the term ''socialist" as follows:  {\it ''Notably, Chavismo is a consciously \textbf{socialist}-feminist practice throughout all of Venezuela. Many communities that before were denied their dignity, have collectively altered their country based on principles of social equity and egalitarianism." \cite{noauthor_defiant_2019}.} 
In contrast,  {\it Red State} uses ''socialist" in a nutpicking way in this example: {\it ``conservative pundits and politicians have painted a devastatingly accurate picture of what happens when a country embraces socialism. Pointing out the dire situation facing the people of Venezuela provided the public with a concrete example of how \textbf{socialist} policies destroy nations."} \cite{charles_selling_nodate}.

\begin{figure}[tb]
\caption{{\bf Abortion bias}}
\hglue-5.89cm\includegraphics[scale=.58]{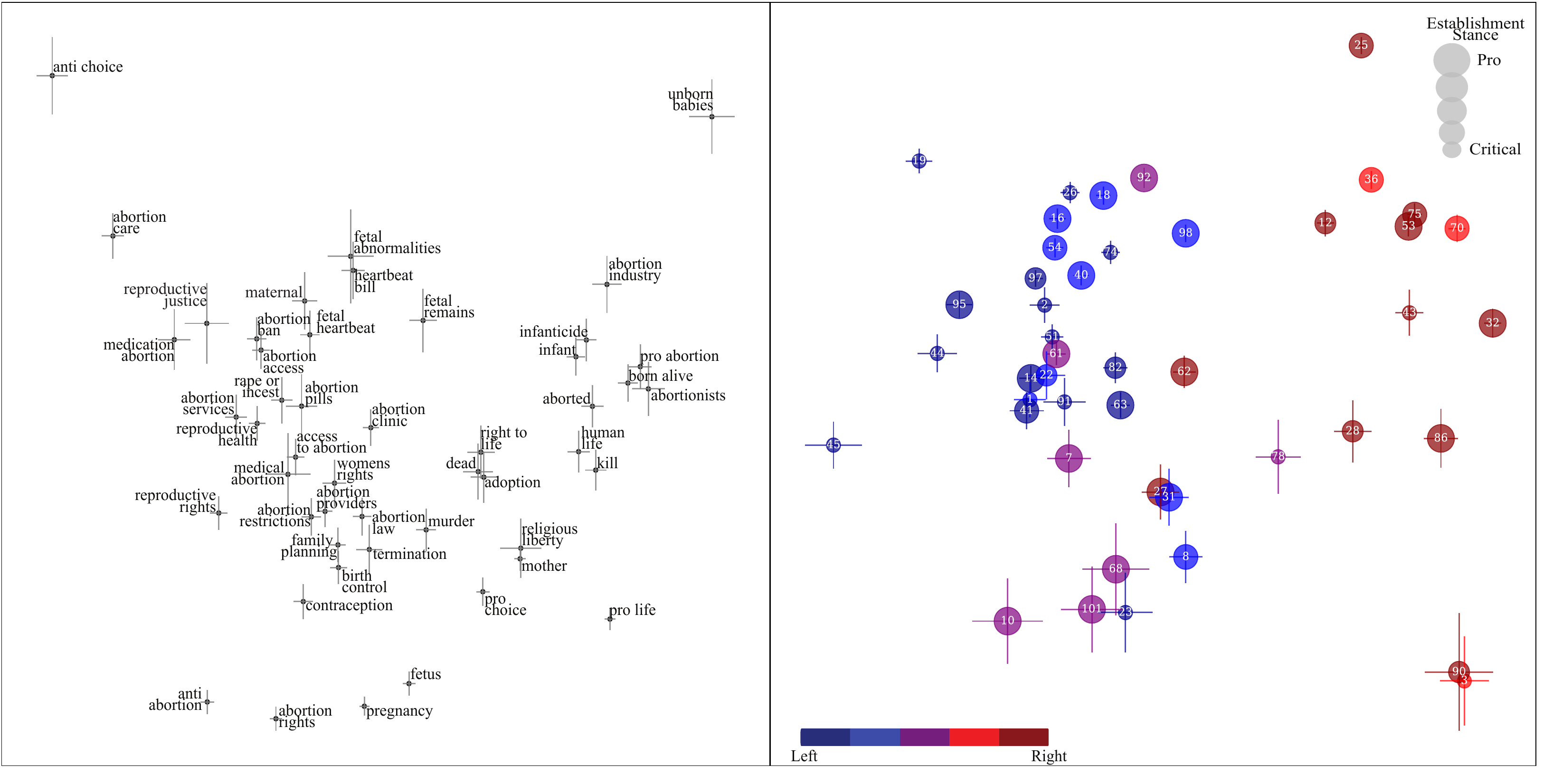}
\vskip-2cm
\label{abortionFig}
\end{figure}

\clearpage

\subsubsection*{Correlated left-right controversies}

Our algorithm auto-discovers bias axes for all the topics we study and, unsurprisingly, many of them reflect a traditional left-right split similar to that revealed by our BLM analysis. For example, \fig{abortionFig} shows that the first principal component (the $x$-axis) for articles on the abortion topic effectively separates newspapers along the left-right axis exploiting relative preferences for terms such as ``fetus''/``unborn babies", ``termination/infanticide" and ``anti choice"/``pro life". 
In addition to valent synonyms, we see that our algorithm detects additional 
bias by differential use of certain phrases lacking obvious counterparts, \eg, 
``reproductive rights" versus ``religious liberty".

\begin{figure}[tb]
\caption{BLM bias (the $x$-axis in \protect\fig{BLMfig}) and abortion bias  (the $x$-axis in \protect\fig{abortionFig}) are seen to be highly correlated. Each dot corresponds to a newspaper (see legend in \protect\fig{legend}).}
\centerline{\includegraphics[scale = .3]{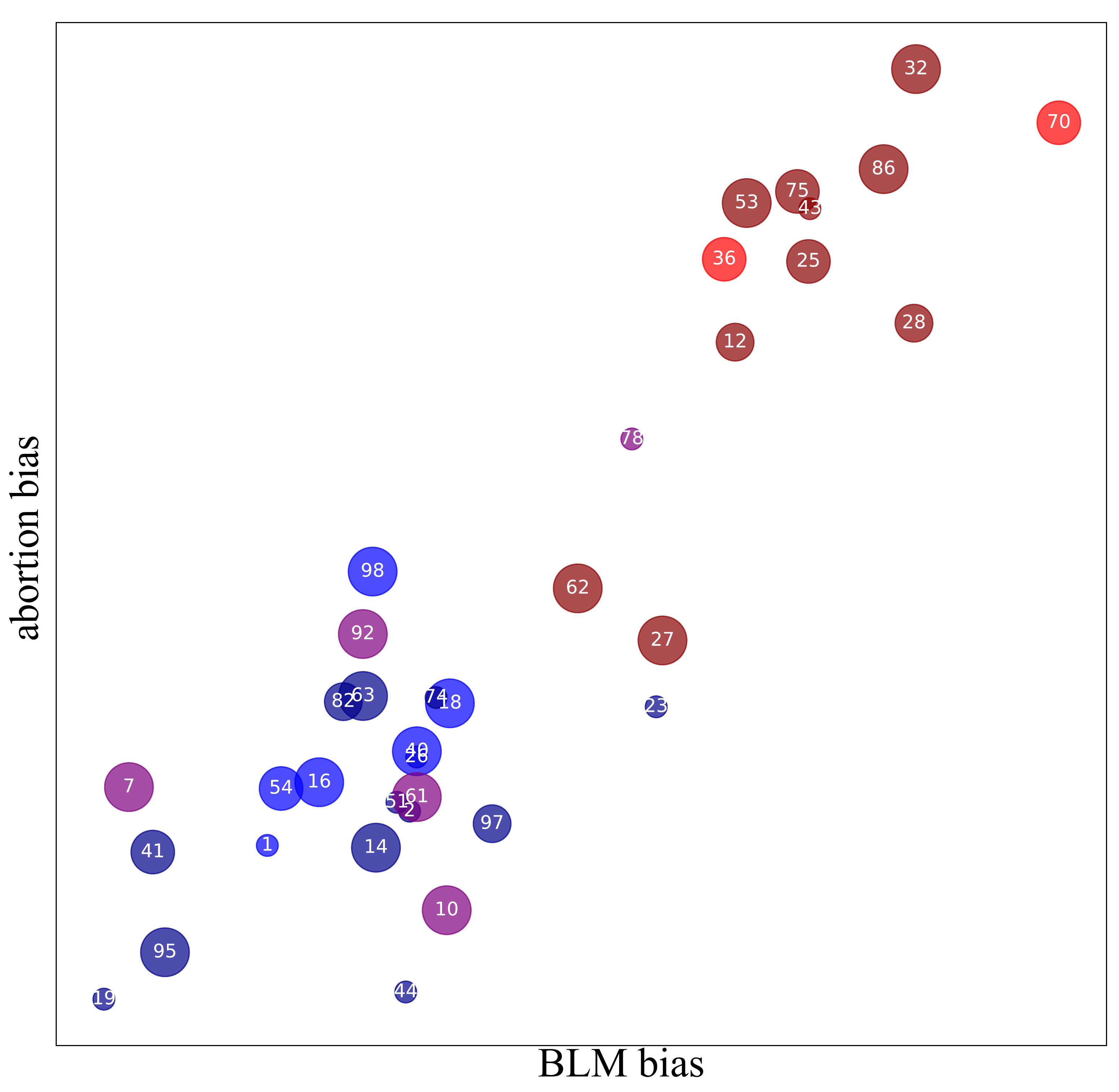}}
\label{BLMabortionCorrFig}
\end{figure}

\Fig{BLMabortionCorrFig} shows that the correlation between BLM bias and abortion bias is very high (the correlation coefficient $r\approx 0.90$). Since these two topics are arguably rather unrelated from a purely intellectual standpoint, their high correlation reflects the well-known bundling of issues in the political system.

\begin{figure}[ht]
\caption{{\bf Bias for tech censorship, US immigration, guns} 
}
\hglue-5cm 
\includegraphics[scale = .82]{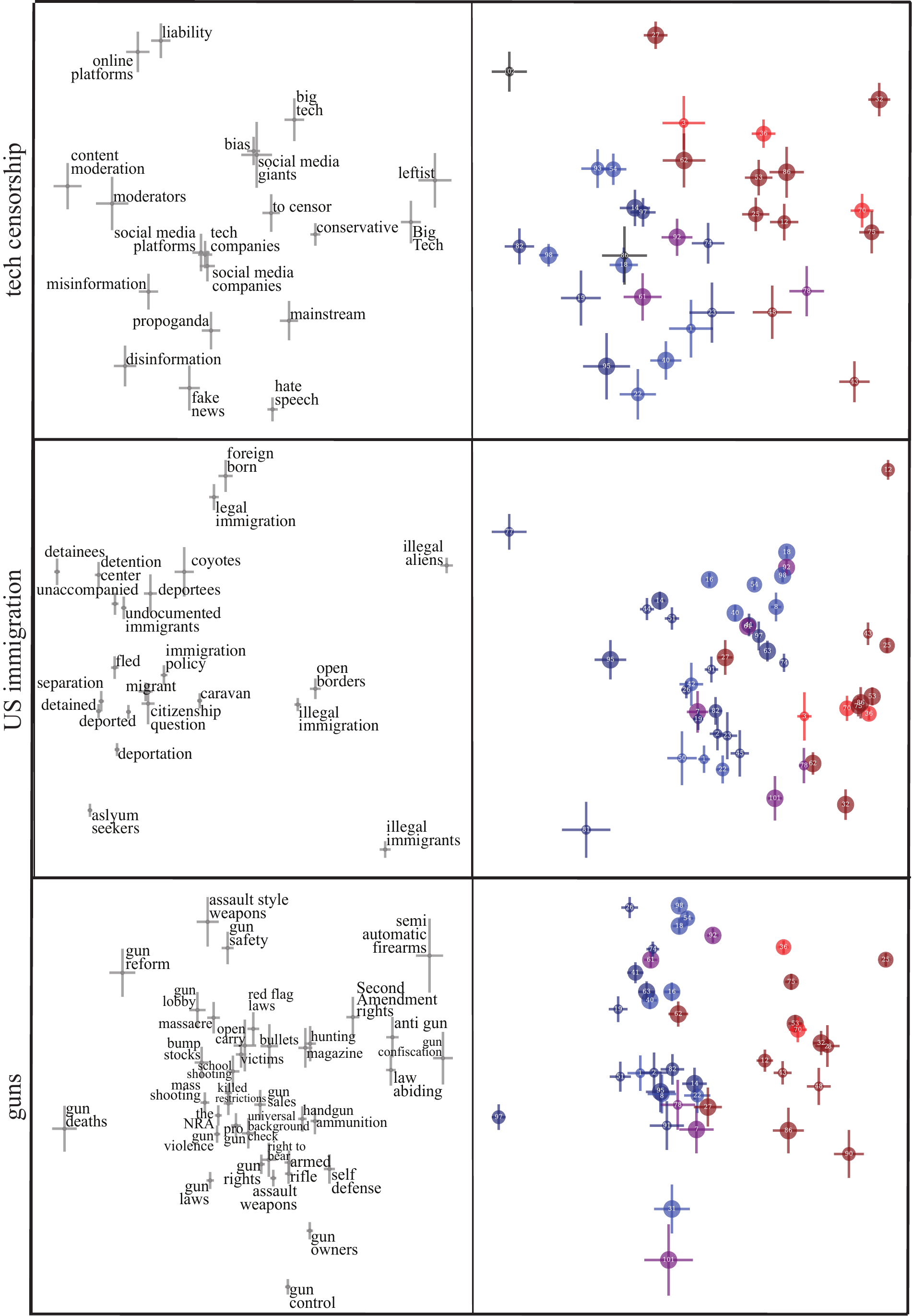}\label{lr_3plots}
\end{figure}

A simple way to auto-identify topics with common bias is to rank topic pairs by their correlation coefficients. 
In this spirit, \tabl{blm_corr} shows the ten topics whose bias is most strongly correlated with BLM bias, together with the corresponding Pearson correlation coefficient $r$ and its standard error
$\Delta r\equiv\sqrt{(1 - r^2)/(n - 2)}$,
where $n$ is the number of newspapers included in its calculation.
The results for three of the most timely top-ranked issues (tech censorship, guns, and US immigration) are shown in \fig{lr_3plots},
again revealing a left-right spectrum of media bias for these topics.

\clearpage

\clearpage

\begin{table}[tb]
\begin{adjustwidth}{-2.25in}{0in} 
\centering
\caption{
{\bf BLM correlation coefficients:}
Topics most correlated with the BLM topic}
\begin{tabular}{|l|l|r|} \hline
\textbf{Topic} & \textbf{Correlation Coefficient} & \textbf{Number of Articles}\\ 
 \thickhline
BLM 1 & $1.00$ & 20044 \\ \hline
abortion 1 & $0.90 \pm 0.05$ & 7541 \\ \hline
tech censorship 1 & $0.85 \pm 0.08$ & 2839 \\ \hline
affirmative action 1 & $0.82 \pm 0.12$ & 8432\\ \hline
US immigration 1 & $0.78 \pm 0.08$ & 23418 \\ \hline
guns 1 & $0.78 \pm  0.08$ & 5444 \\\hline
Russia 1 & $0.71 \pm 0.09$ & 138479 \\ \hline
universities 1 & $0.68 \pm 0.10$ & 8432 \\ \hline
sexual harassment 1 & $0.62\pm 0.10$ & 4521 \\ \hline
Israel 2 & $0.57 \pm 0.11$ & 43406 \\ \hline
church state 1 & $0.55 \pm 0.16$ & 22124 \\ \hline
\end{tabular}
\begin{flushleft}
\end{flushleft}
\label{blm_corr}
\end{adjustwidth}
\end{table}

\subsection*{Establishment bias}
\begin{figure}[tb]
\caption{{\bf Pro-Critical Bias Spectrum: Military spending}
Generalized principal components for military spending.}
\hglue-5.89cm\includegraphics[scale=.58]{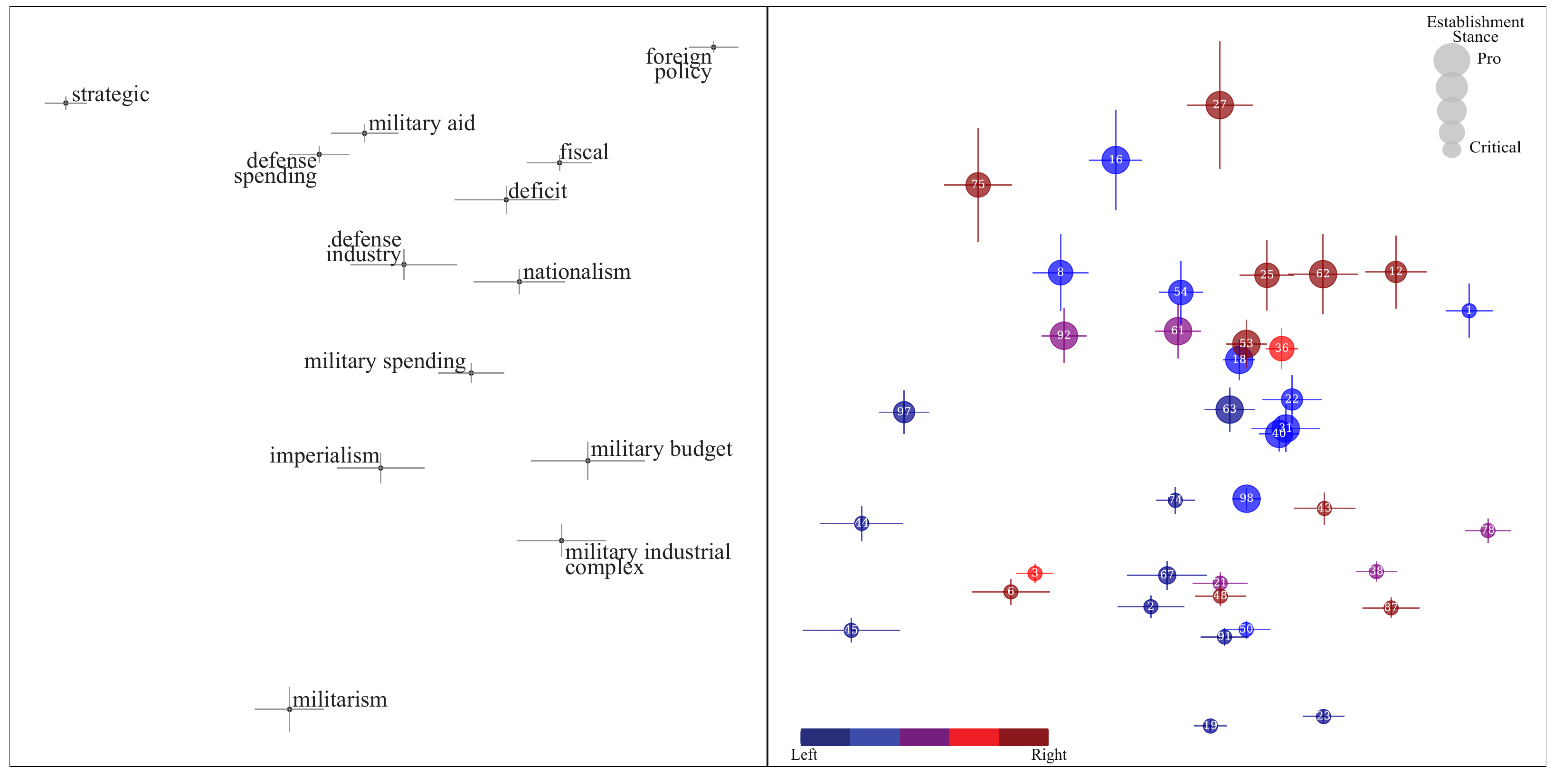}
\label{military_spending}
\end{figure}

The figures above show that although the left-right media axis explains some of the variation among newspapers, it does not explain everything. 
Figure \ref{military_spending} shows a striking example of this for the military spending topic.
As opposed to the previous bias plots, the dots are no longer clearly separated by color (corresponding to left-right stance). 
Indeed, left-leaning CNN (18) is seen right next to right-leaning National Review (53) and Fox News (36).
Instead, the dots are seen to be vertically separated by {\it size}, corresponding to establishment stance. 
In other words, we have auto-identified a second bias dimension, here ranging vertically from establishment-critical (bottom) to pro-establismnent (top) bias.

Just like left-right bias, establishment bias manifests as differential phrase use.
For example, as seen Table \ref{ms_counts}. the phrase ``military industrial complex" is used more frequently in newspapers classified as establishment-critical, such as {\it Canary} and {\it American Conservative}, but is rarely, if ever, used by mainstream, pro establishment outlets such as Fox or CNN, which instead prefer phrases such as ``defense industry".

\begin{figure}[tb]
\caption{{\bf Correlation between military spending bias and Venezuela bias}}
\includegraphics[scale = .3]{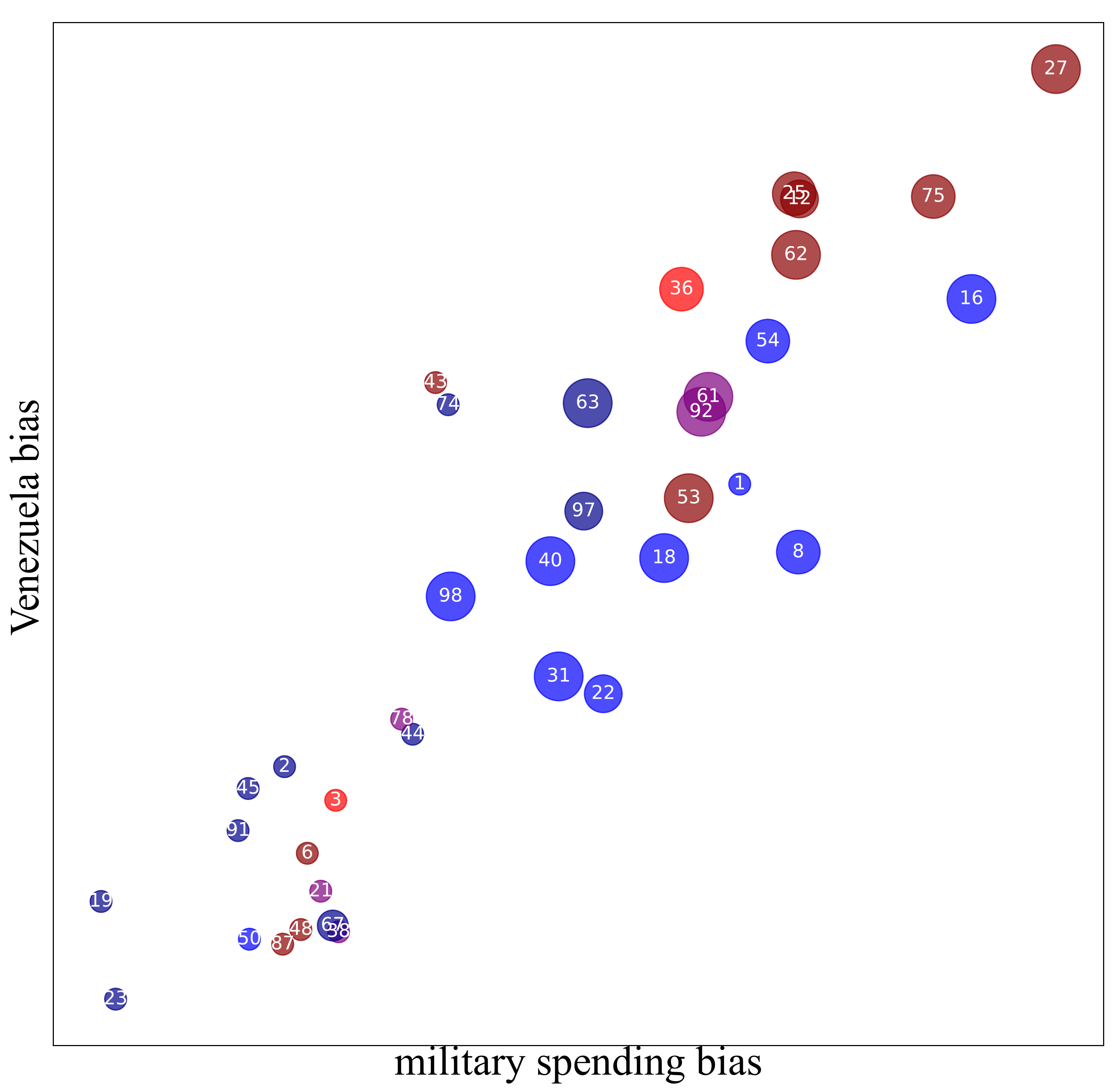}
\label{militaryVen}
\end{figure}

\begin{figure}[tb]
\caption{{\bf Establishment bias for Venezuela}\\}
\hglue-5.89cm\includegraphics[scale=.58]{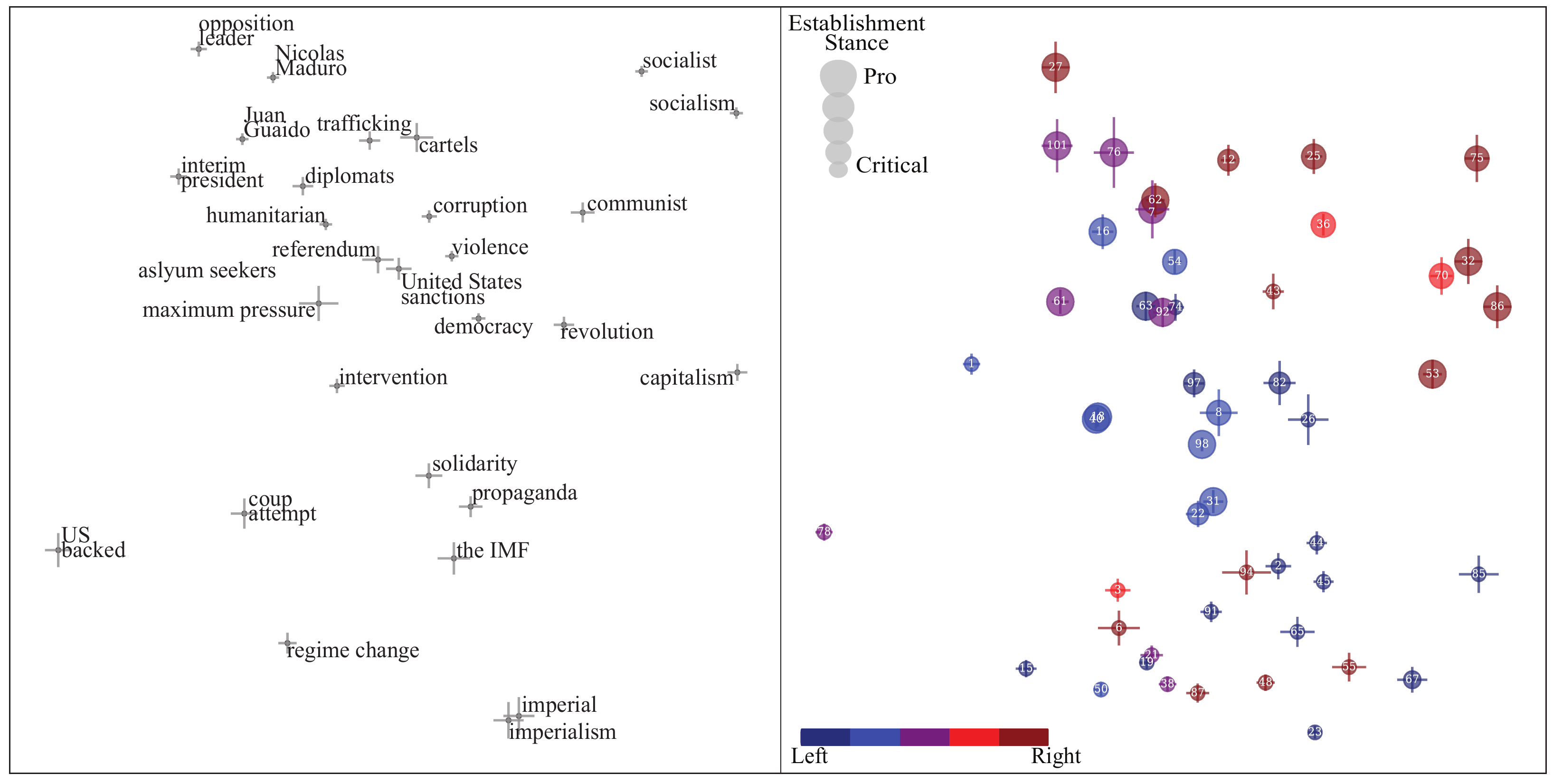}
\label{ven}
\end{figure}

We find that the military spending topic, much like the BLM topic, is highly correlated with other topics included in the study. This is clearly seen in Figure \Fig{militaryVen}, which plots the pro-critical generalized principal components of the military spending topic and the Venezuela topic. 
A closer look at the Venezuela topic in \Fig{ven} reveals a establishment bias similar to that seen in \fig{military_spending}. We see that, while establishment-critical papers frequently use phrases such as ``imperialism" and ``regime change", pro-establishment newspapers prefer phrases such as ``socialism" and ``interim president". 
This figure reveals that the Venezuela topic engenders both establishment bias (the vertical axis) and also a smaller but non-negligible left-right bias (the horizontal axis).

\begin{table}[tb]
\begin{adjustwidth}{-2.25in}{0in} 
\centering
\caption{
{\bf Military spending phrase usage per article}}
\begin{tabular}{|l|r|r|r|r|} \hline
\textbf{Phrase} & \textbf{Canary} & \textbf{American Conservative} & \textbf{Fox} &  \textbf{CNN} \\
 \thickhline
militarism &  $0.42 \pm 0.11$ & $0.05 \pm 0.02$ &  $0.00 \pm 0.00$ &  $0.00 \pm 0.00$ \\ \hline
military industrial complex &  $0.92 \pm 0.16$ & $0.27 \pm 0.04$ &  $0.00 \pm 0.00$ &  $0.00 \pm 0.00$ \\ \hline
defense spending & $0.00 \pm 0.00$ & $0.54 \pm 0.06$ &  $0.24 \pm  0.02$&  $0.51 \pm 0.02$ \\ \hline
military aid & $0.00 \pm 0.00$ & $0.00 \pm 0.00$ & $0.03 \pm 0.01$ & $0.06 \pm 0.01$\\ \hline
\end{tabular}
\label{ms_counts}
\end{adjustwidth}
\end{table}

\clearpage

\begin{table}[tb]
\begin{adjustwidth}{-2.25in}{0in} 
\centering
\caption{
{\bf Topics whose bias is most correlated with 
military spending bias}}
\begin{tabular}{|l|c|r|} \hline
\textbf{Topic} & \textbf{Correlation Coefficient} & \textbf{Number of Articles}\\ 
 \thickhline
military spending 2 & $1.00 \pm 0.00 $& 45802 \\ \hline
Venezuela 2 & $0.90 \pm 0.05 $& 17142 \\ \hline
public finance 2 & $0.84 \pm 0.07$ &  79076\\ \hline
human rights 2 & $0.76 \pm 0.08$ & 7623 \\ \hline
nuclear weapons 2 & $0.75 \pm 0.09$ & 45802 \\ \hline
Yemen 2 & $0.75 \pm 0.08$ & 12835 \\ \hline
Israel 2 & $0.69 \pm 0.09$ & 43406 \\ \hline
Palestine 2 & $0.68 \pm 0.10$ & 5461 \\ \hline
private finance 2 & $0.68 \pm 0.10$ & 79076 \\\hline
prisons 2 & $0.64 \pm 0.10$ & 79076 \\\hline
private finance 1 & $0.64 \pm 0.11$ & 79076 \\\hline
police 2 & $0.61 \pm 0.11$ & 20044 \\ \hline
\end{tabular}
\label{ms_corr}
\end{adjustwidth}
\end{table}

\clearpage

\begin{figure}[tb]
\caption{{\bf Establishment bias for nuclear weapons, Yemen, police}}
\hglue-6cm \includegraphics[scale = .82]{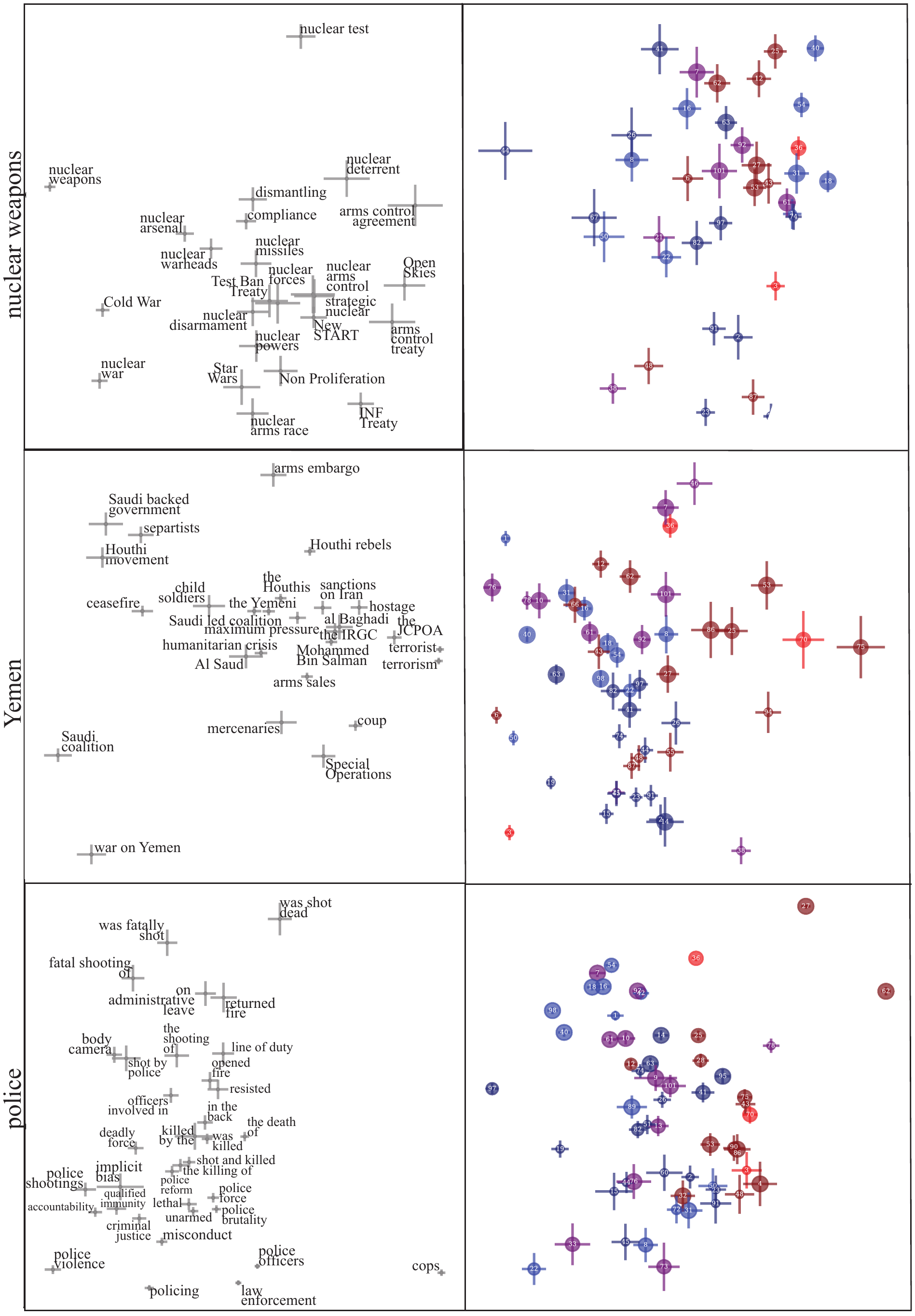}
\label{pc_3plots}
\end{figure}

\clearpage

\begin{figure}[tb]
\caption{{\bf Valent synonyms reflecting establishment bias}
}
\includegraphics[scale = .6]{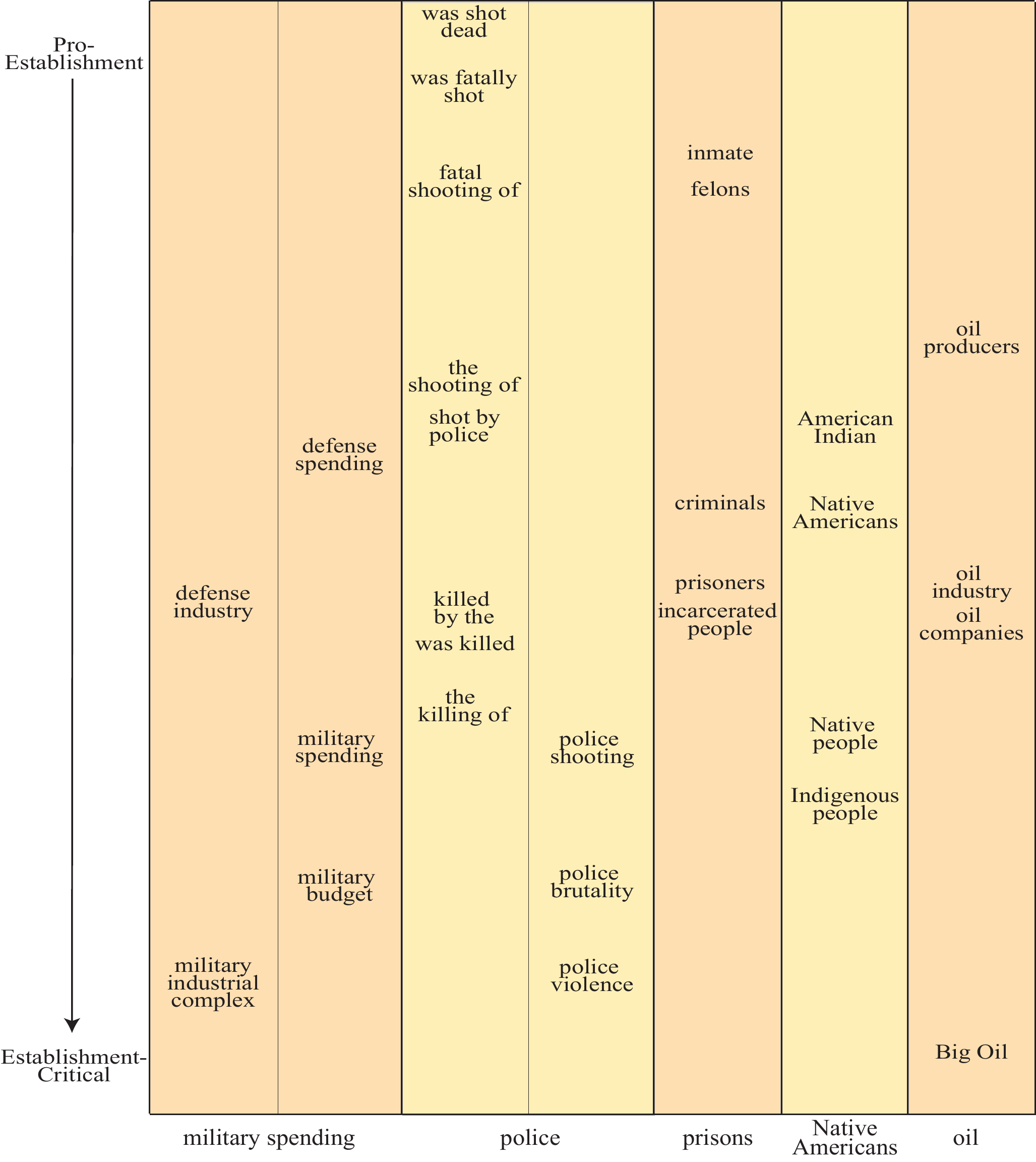}
\vskip-2mm
\label{pro_crit-phrases}
\end{figure}

To identify additional topics with establishment bias, we again compute 
correlation coefficients between generalized principal 
components---this time with the vertical component for
military spending. 
Table \ref{ms_corr} shows the ten most correlated topics,
revealing a list quite different from the 
left-right-biased topics from \Tabl{blm_corr}. 
Nuclear weapons, 
Yemen, and police, three timely examples from this list, are shown in Figure 
\ref{pc_3plots}. 
Here the left panels illustrate how usage of certain phrases reflects 
establishment bias separation. In articles about nuclear weapons,
the terms ``nuclear arms race" and ``nuclear war" are seen to appear 
preferentially in establishment-critical newspapers, while ``nuclear test" and 
``nuclear deterrent" are preferred by pro-establishment papers.
In articles about Yemen, the phrase ``war on Yemen", suggesting a clear cause, 
is seen to signal an establishment-critical stance, while
``humanitarian crisis", not implying a cause, signals pro-establismnent stance.
For articles about police, grammatical choices in the coverage of police 
shootings is seen to be highly predictive of establishment stance:
establishment-critical papers use passive voice (\eg, ``was shot dead") less 
than pro-establishment papers, and when they do, they prefer the verb ``killed" 
over ``shot". Such news bias through use of passive voice was explored in detail in \cite{herman2010manufacturing}.
\Fig{pro_crit-phrases} illustrates such use of the passive voice 
and valent synonyms across establishment topics.

\clearpage

\begin{figure}[t]
	\caption{{\bf Bias correlation Matrix}
	}
	\hglue-1.4cm 
	\includegraphics[scale = .35]{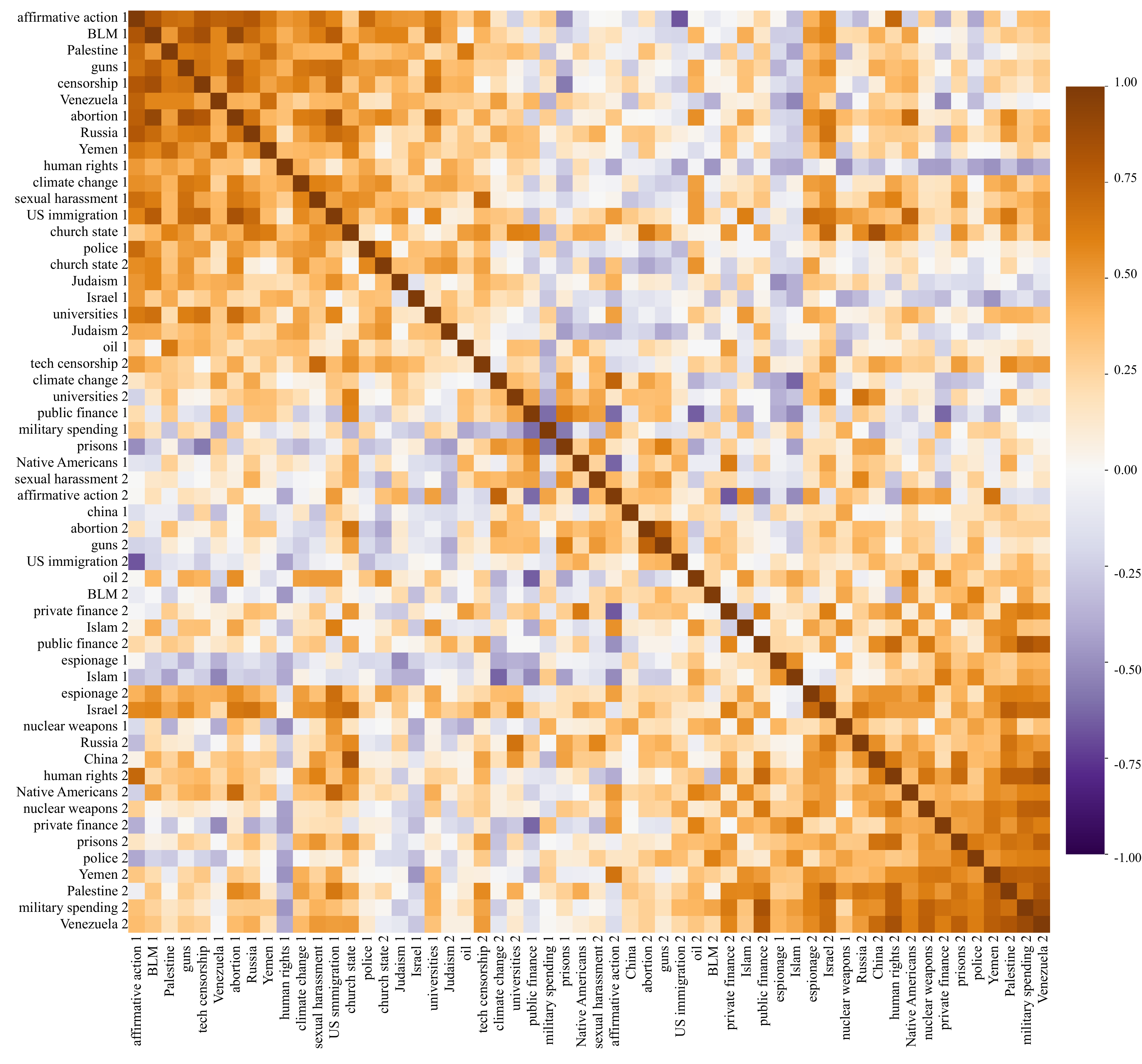}
	\vskip-2mm
	\label{corr_matrix}
\end{figure}

\section*{Machine learning the media bias landscape}
\begin{figure}[tb] 
\caption{{\bf Spectral clustering of topics by their media bias characteristics} as explained in the text. The bars represent 1 standard deviation Jackknife error bars.}\includegraphics[scale =.57]{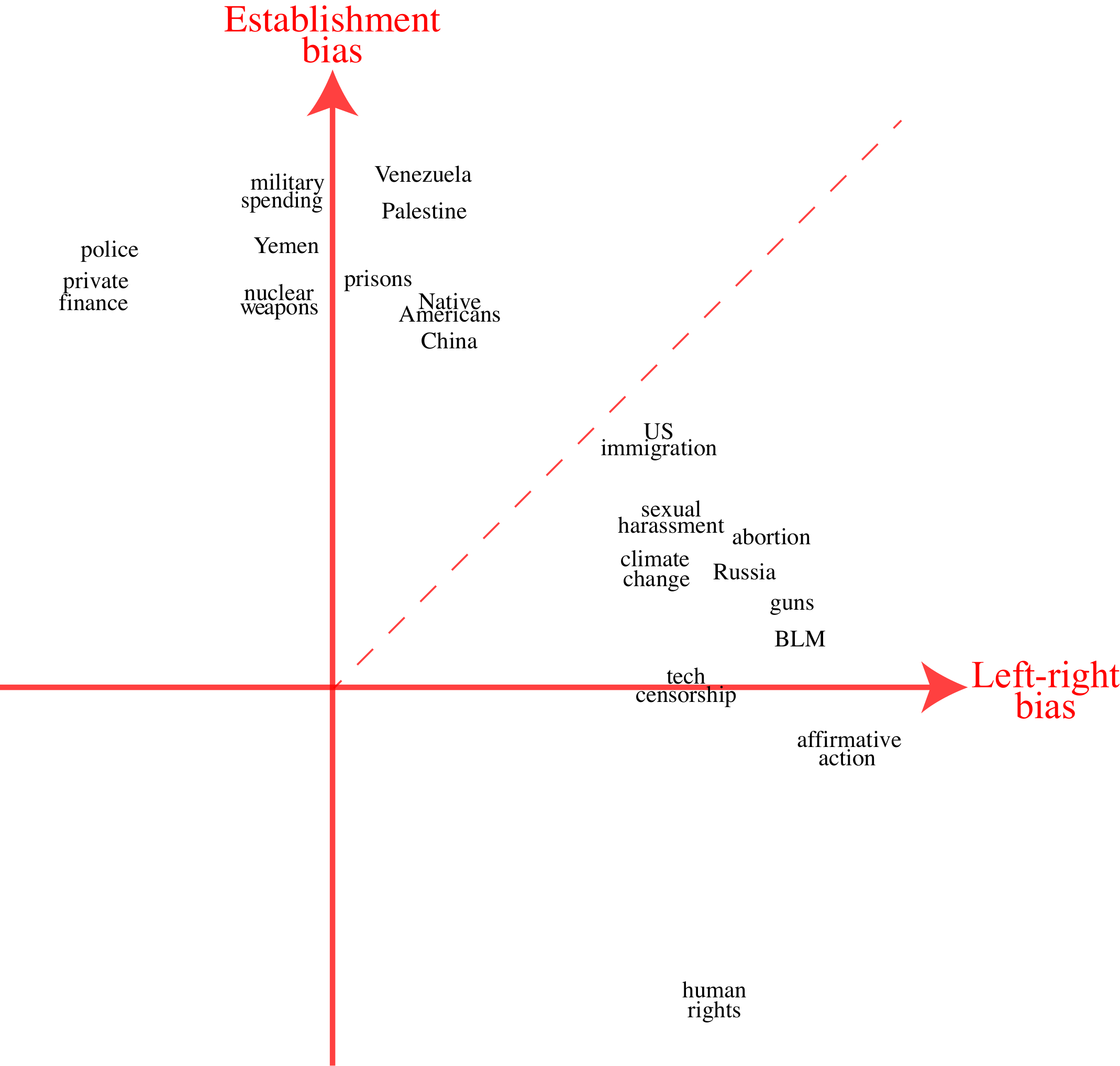}
 \label{dartboardFig}
\end{figure}

Throughout this paper, we have aspired to measure media bias in a purely 
data-driven way, so that the data can speak for itself without human 
interpretation. In this spirit, we will now eliminate the manual elements from our above bias landscape exploration
(our selection of the two rather uncorrelated topics {\it BLM} and {\it military spending} and the topics most correlated with them).
Our starting point is the $56\times 56$ correlation matrix $\R$ for the generalized principal 
components of all our analyzed topics, shown in \fig{corr_matrix}. 
Notation such as ``BLM 1" and ``BLM 2" reflects the fact that we have two generalized principal 
components corresponding to each topic (the two axes of the right panel of \fig{BLMfig}, say).
Our core idea is to use the standard technique of spectral clustering \cite{ng2002spectral} to identify which topics exhibit similar bias, 
using their bias correlation from \fig{corr_matrix} as measure of similarity.
We start by performing an eigendecomposition
\beq{SpectralDecompositionEq}
R_{ij} = \sum_i\lambda_k E_{ik} E_{jk}
\eeq
of the correlation matrix $\R$, where $\lambda_i$ are the eigenvalues and the columns of the matrix $\E$ are the eigenvectors.
\Fig{dartboardFig} illustrates the first two eigencomponents, with the point corresponding to the $\kth$ topic plotted at coordinates $(E_{1k},E_{2k})$.
To reduce clutter, we show the ten components with the largest $|E_{1k}|$ and the ten with the largest $|E_{2k}|$, retaining only the largest component for each topic. For better intuition, the figure has been rotated by $45^\circ$, since if two internally correlated clusters are also correlated with each other, this will tend to line up the clusters with the coordinate axes. If needed, we also flip the sign of any axis whose data is mainly on the negative side and flip the 1/2 numbering to reflect cluster membership as described in the Supporting Information.

We can think of \fig{dartboardFig} as mapping all topics into a 2-dimensional media bias landscape.
The figure reveals a clear separation of the topics into two clusters based on their media bias characteristics. 
A closer look at the membership of these two clusters suggests 
interpreting the $x$-axis as left-right bias and the $y$-axis as establishment bias.
We therefore auto-assign each topic to one of the two clusters based on whether it falls closer to the $x$-axis or the $y$-axis
(based whether $|E_{1k}|>|E_{2k}|$ or not, in our case corresponding to which side of the dashed diagonal line the topic falls).
We then sort the topics on a spectrum from most left-right-biased to most establishment-biased:
the left-right topics are sorted by decreasing $x$-coordinate and followed by the establishment topics sorted by increasing $y$-coordinate.
When ordered like this, the two topic clusters become visually evident even in the correlation matrix $\R$ upon which our clustering analysis was based:
\fig{corr_matrix} shows two clearly visible blocks of highly correlated topics – both the left-right block in the upper left corner and the establishment block in the lower right. 

 \clearpage

\begin{figure}[tb] 
\caption{{\bf Media  bias landscape:}
Our method locates newspapers into this two-dimensional media bias landscape 
based only on how frequently they use certain discriminative phrases, with no human input regarding what constitutes bias. The colors and sizes of the dots were predetermined by external assessments and thus in no way influenced by our data. 
The positions of the dots thus suggest that the two dimensions can be 
interpreted as the traditional left-right bias axis and establishment bias, respectively. 
}\label{MediaLandscapeFig}\hglue-6cm \includegraphics[scale =.33]{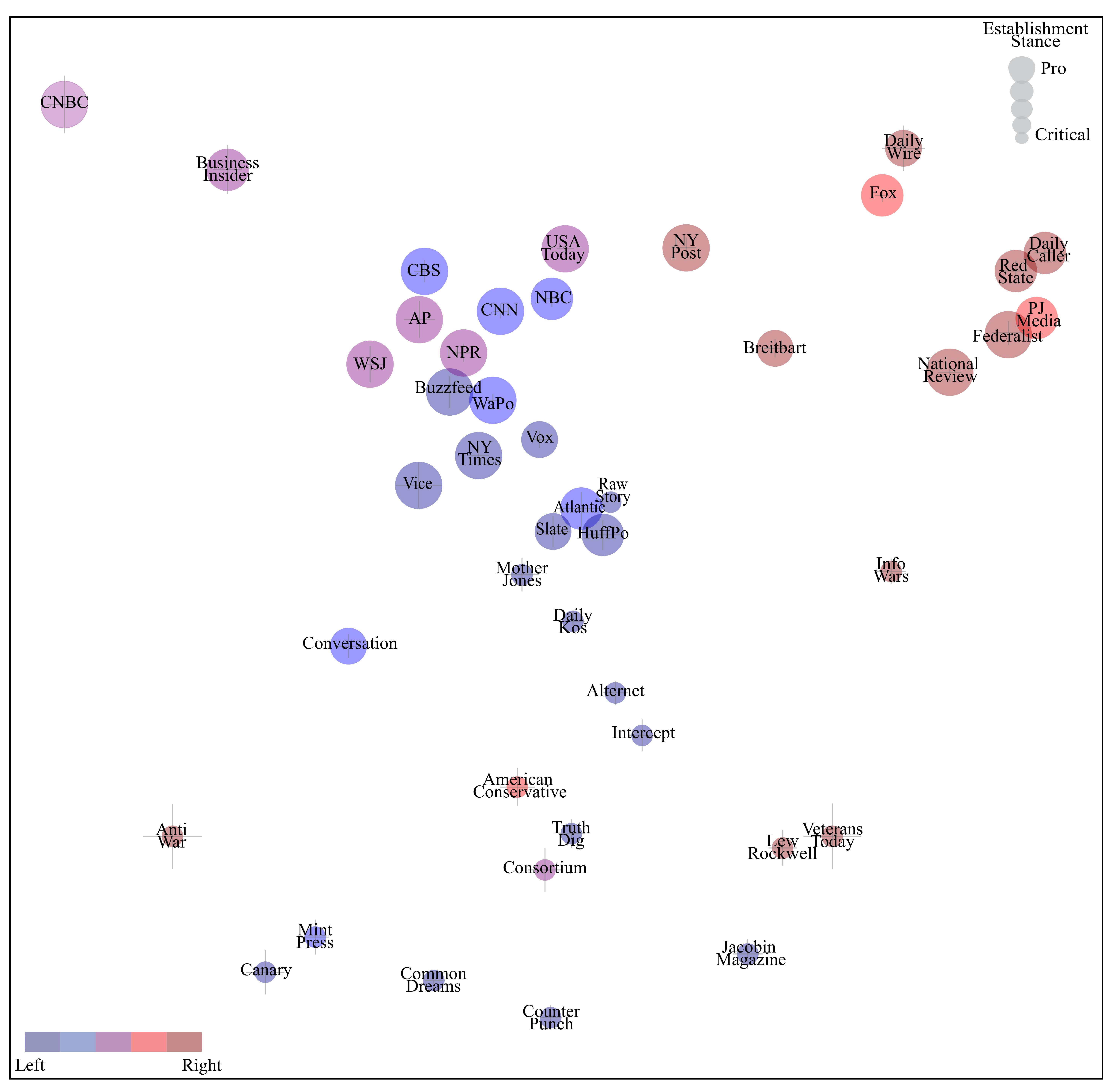} 
\end{figure}

\clearpage

Above, the newspapers were mapped onto a separate bias plane for each of many different topics. 
We normalize each such media plot, \eg, the left panel of \fig{BLMfig}, such that the dots have zero mean and unit variance both horizontally and vertically. 
We then unify all these plots into a single media bias landscape plot in \fig{MediaLandscapeFig} by taking weighted averages of these many topic plots, weighting both by topic relevance and inverse variance.
Specifically,  for each topic bias, we assign two relevance weights corresponding to the absolute value of its $x$- and $y$-coordinates in \fig{dartboardFig}, reflecting its relevance to left-right and establishment bias, respectively. These weights can be found in the Supporting Information.
For example, to compute the $x$-coordinate of a newspaper in \fig{MediaLandscapeFig}, we
simply take a weighted average of its generalized principal components for all topics, weighted both by the left-right relevance of that topic and by the inverse square of the error bar.

\Fig{MediaLandscapeFig} can be viewed as the capstone plot for this paper, unifying  information from all our topic-specific bias analyses.
It reveals fairly good agreement with our the external human-judgement-based bias classifications reflected by the colors and sizes of the dots: it shows a separation between blueish does on on the left and reddish ones on the right, as well as a separation between larger (pro-establishment) dots toward the top and smaller (establishment-critical) ones toward the bottom.

Closer inspection of \fig{MediaLandscapeFig} also reveal some notable exceptions that deserve further scrutiny.
As mentioned, the ``nutpicking" poses a challenge for our method. An obvious example is {\it Jacobin Magazine}, a self-proclaimed socialist newspaper \cite{jacobin} that \fig{MediaLandscapeFig} classifies as right-leaning because of its heavy use of the phrase ``socialism" approvingly while it is mainly used pejoratively by right-leaning media.
Nutpicking may also help explain why 
\fig{MediaLandscapeFig} shows some more extreme newspapers closer to the center than more moderate ones (according to the human-judgement-based classification from AllSides \cite{noauthor_allsides_nodate}).
For example, AllSides rates {\it Breitbart} as further right than {\it Fox}, and uses the phrase ``defund the police" more often than {\it Fox} --- presumably to criticize or mock it, thus getting pulled to the left in \fig{MediaLandscapeFig} towards left-leaning newspapers who use the phrase approvingly.
One might expect nutpicking to be more common on the extremes of the political spectrum, in which case our method would push these newspapers toward the center.
\fig{MediaLandscapeFig} also shows examples where our method might be outperforming the human-judgement-based classification from AllSides \cite{noauthor_allsides_nodate}). For example, 
\cite{noauthor_allsides_nodate} labels 
{\it Anti War} as ``right" while our method classifies it as left, in better agreement with its online mission statement. 

Our analysis also offers more nuance than a single left-right bias-score: for example, our preceding plots show that 
{\it American Conservative} is clearly right-leaning on social issues such as abortion and immigration, while clearly left on issues involving foreign intervention, averaging out to a rather neutral placement in \fig{MediaLandscapeFig}.

\section*{Conclusions}

We have presented an automated method for measuring media bias. 
It first auto-discovers the phrases whose frequencies contain the most information about what newspapers published them, and then uses observed frequencies of these phrases to map newspapers into a two-dimensional media bias landscape. 
We have roughly a million articles from about a hundred newspapers for bias in dozens of news topics, producing a 
a data-driven bias classification in good agreement with prior classifications based on human judgement. One dimension can be interpreted as traditional left-right bias, the other as establishment bias.

Our method leaves much room for improvement, and we will now mention three examples.
First, we saw how the popular practice of {\bf nut-picking} can cause problems for our analysis by the same phrase being used with positive or negative connotations depending on context. This could be mitigated by excluding such bi-valent phrases from the analysis, either manually or with better machine learning.

Second, {\bf topic bias} can cause challenges for our method, by separating newspapers by their topic focus (say business versus sports) in a way that obscures political bias. As described above, we attempted to minimize this problem by splitting overly broad topics into narrower ones, but this process should be improved and ideally automated.

Third, although our method is almost fully automated, a {\bf manual screening step} remains whereby auto-selected phrases are discarded if they lack sufficient relevance, uniqueness or specificity. Although this involves only the {\it selection} of phrases (machine-learning features), not their {\it interpretation}, it is worthwhile exploring whether this screening can be further (or completely) automated, ideally making our method completely free of manual steps and associated potential for human errors.

As datasets and analysis methods continue to improve, the quality of automated news bias classification should get ever better, enabling more level-headed scientific discussion of this important phenomenon.
We therefore hope that automated new bias detection can help make discussions of media bias less politicized than the media being discussed.

\bigskip
{\bf  Acknowledgements:}
The authors wish to thank 
Rahul Bhargava,
Meia Chita-Tegmark,
Haimoshri Das,
Emily Fan,
Jamie Fu,
Finnian Jacobson-Schulte,
Dianbo Liu,
Jianna Liu, 
Mindy Long,
Hal Roberts,
Khaled Shehada,
Arun Wongprommoon, 
and
Ethan Zuckerman,
for helpful comments and support during the launch phase of this project.
This work was supported by the Foundational Questions Institute, 
the Rothberg Family Fund for Cognitive Science and IAIFI through NSF grant PHY-2019786.

\bibliography{wordbias}

\clearpage
\section*{Supporting Information}\label{S1-Appendix}

In this section, we include supplementary technical details in the form of additional generalized eigenvalue plots and topic relevance weights.

\begin{figure}[h] \hglue-4cm \caption{\bf{Affirmative
 action bias}}\includegraphics[scale =.4]{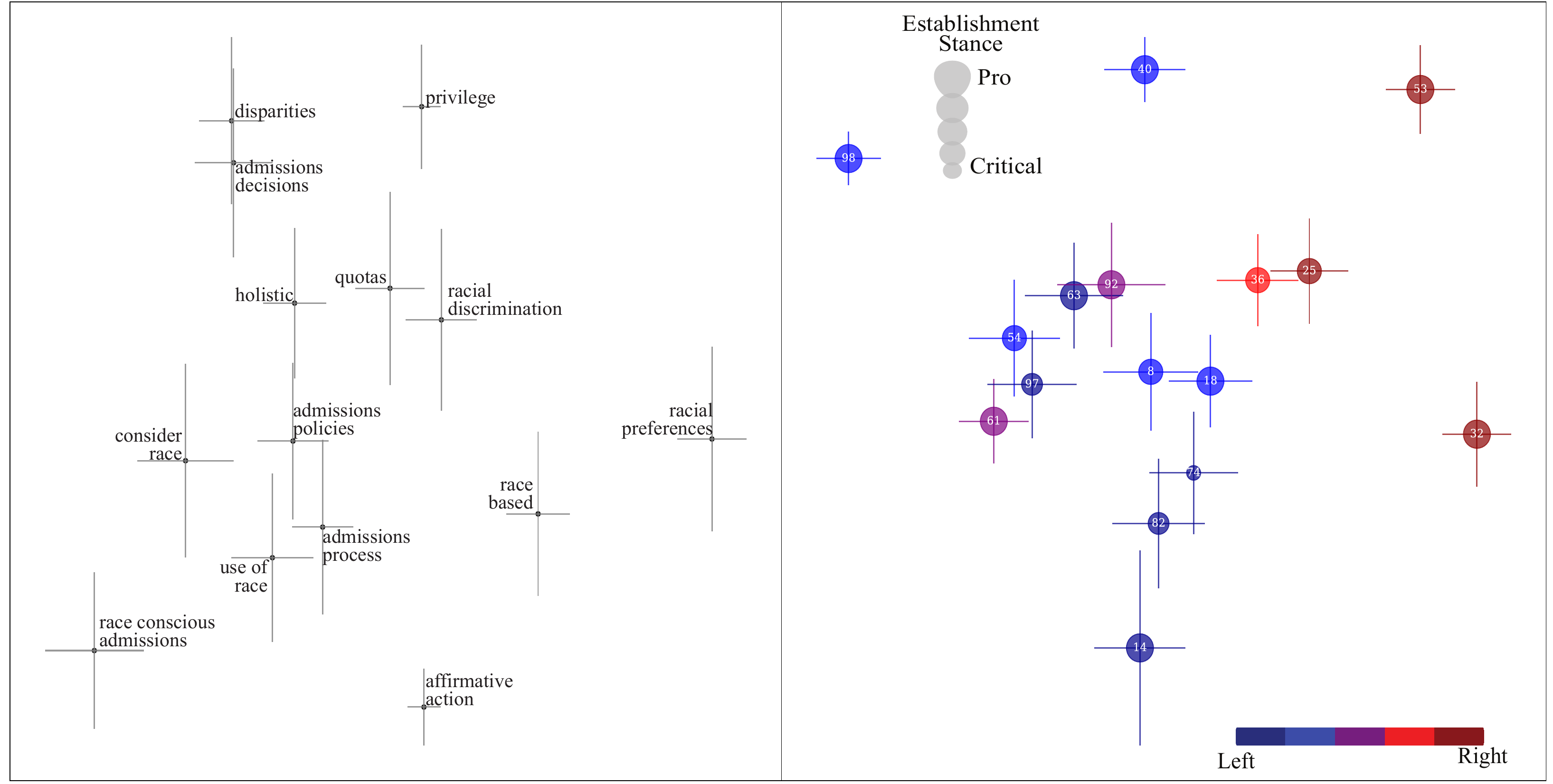}\label{affirmative_action}
\end{figure}

\begin{figure}[h] \hglue-4cm \caption{\bf Palestine bias}\includegraphics[scale =.4]{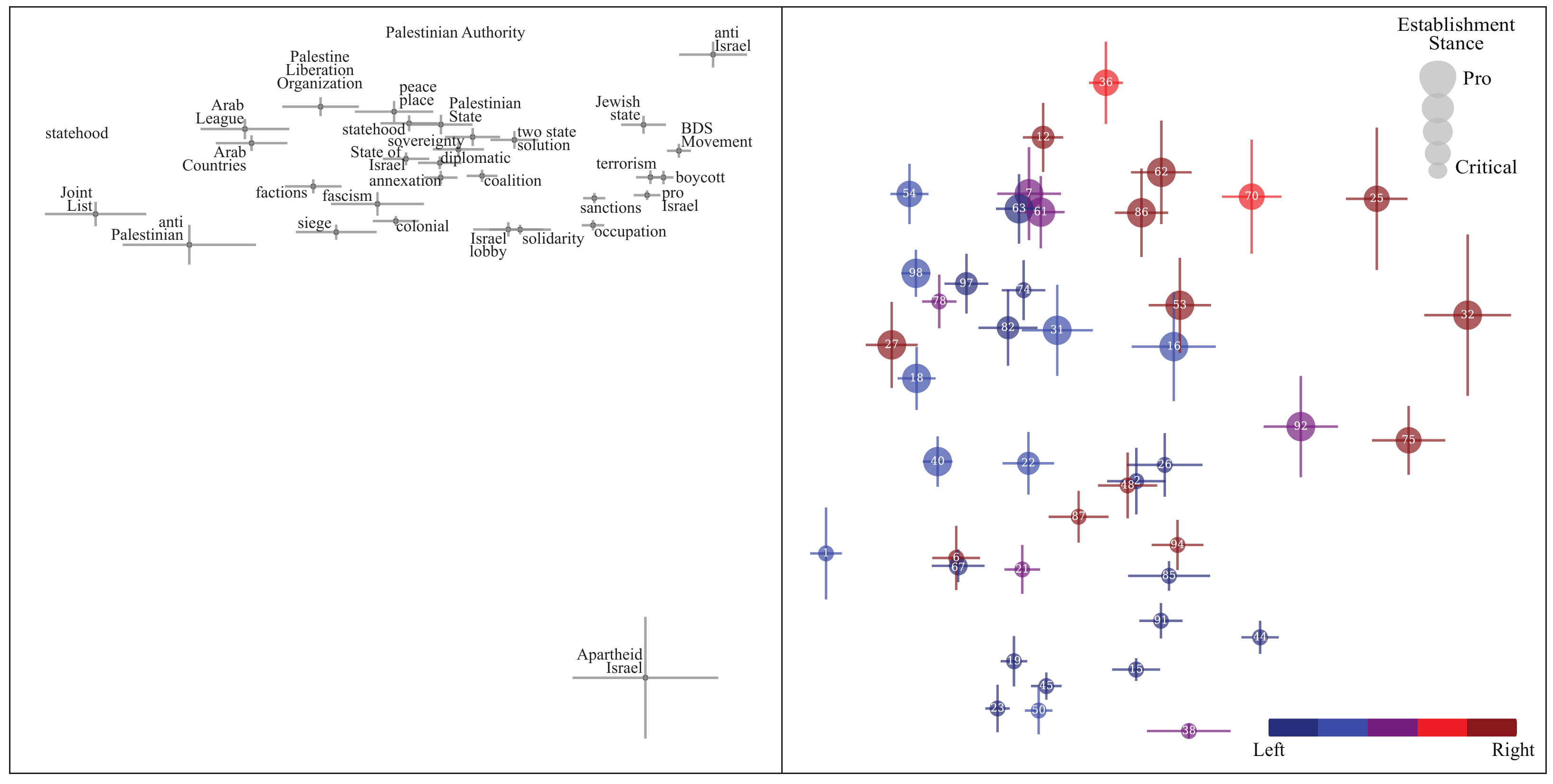}\label{PSE}
\end{figure}

\begin{figure}[h] \hglue-4cm \caption{\bf Public finance bias}\includegraphics[scale =.4]{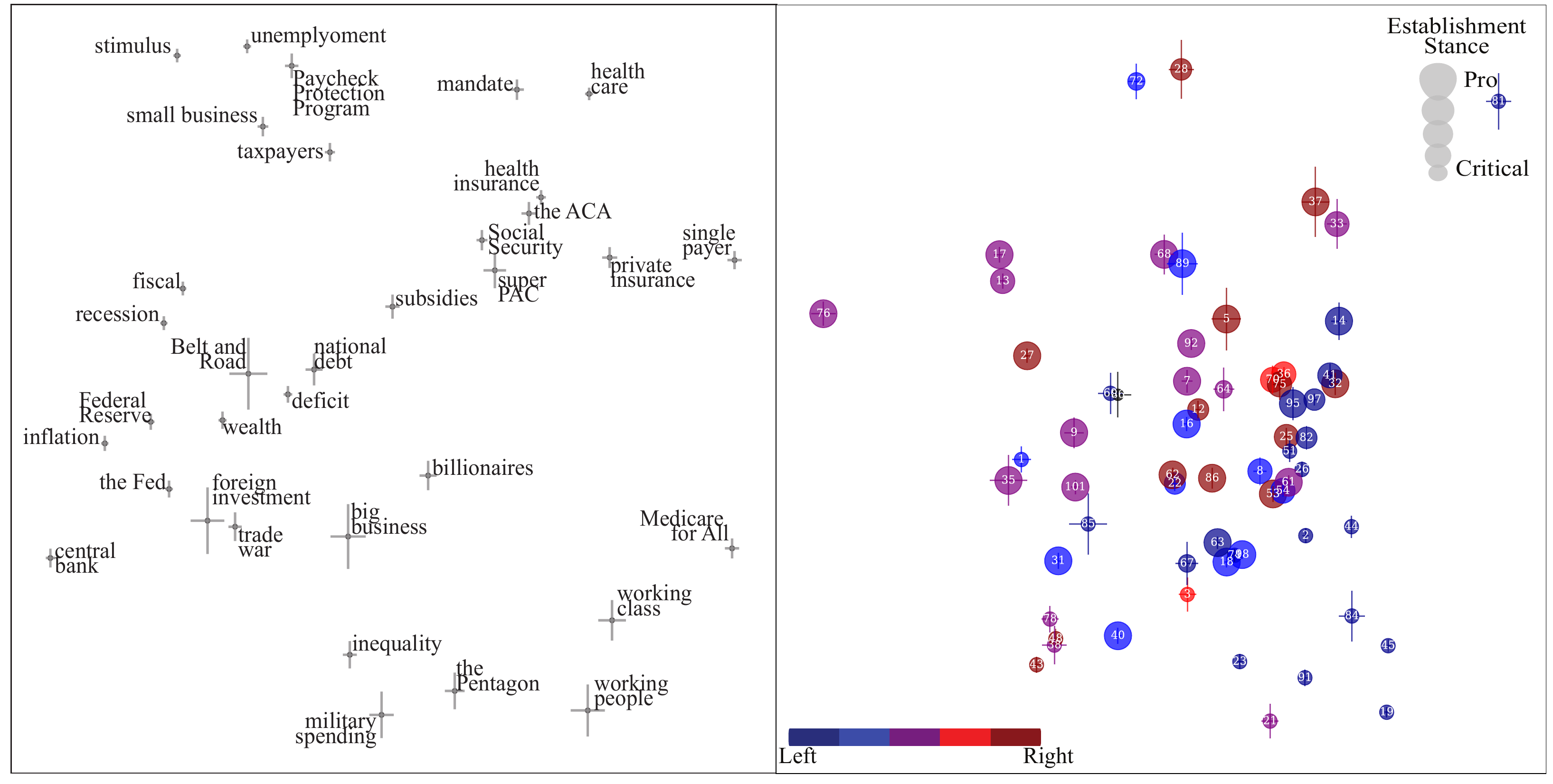}\label{publicfinance}
\end{figure}

\begin{figure}[h] \hglue-4cm \caption{\bf Human Rights bias}\includegraphics[scale =.4]{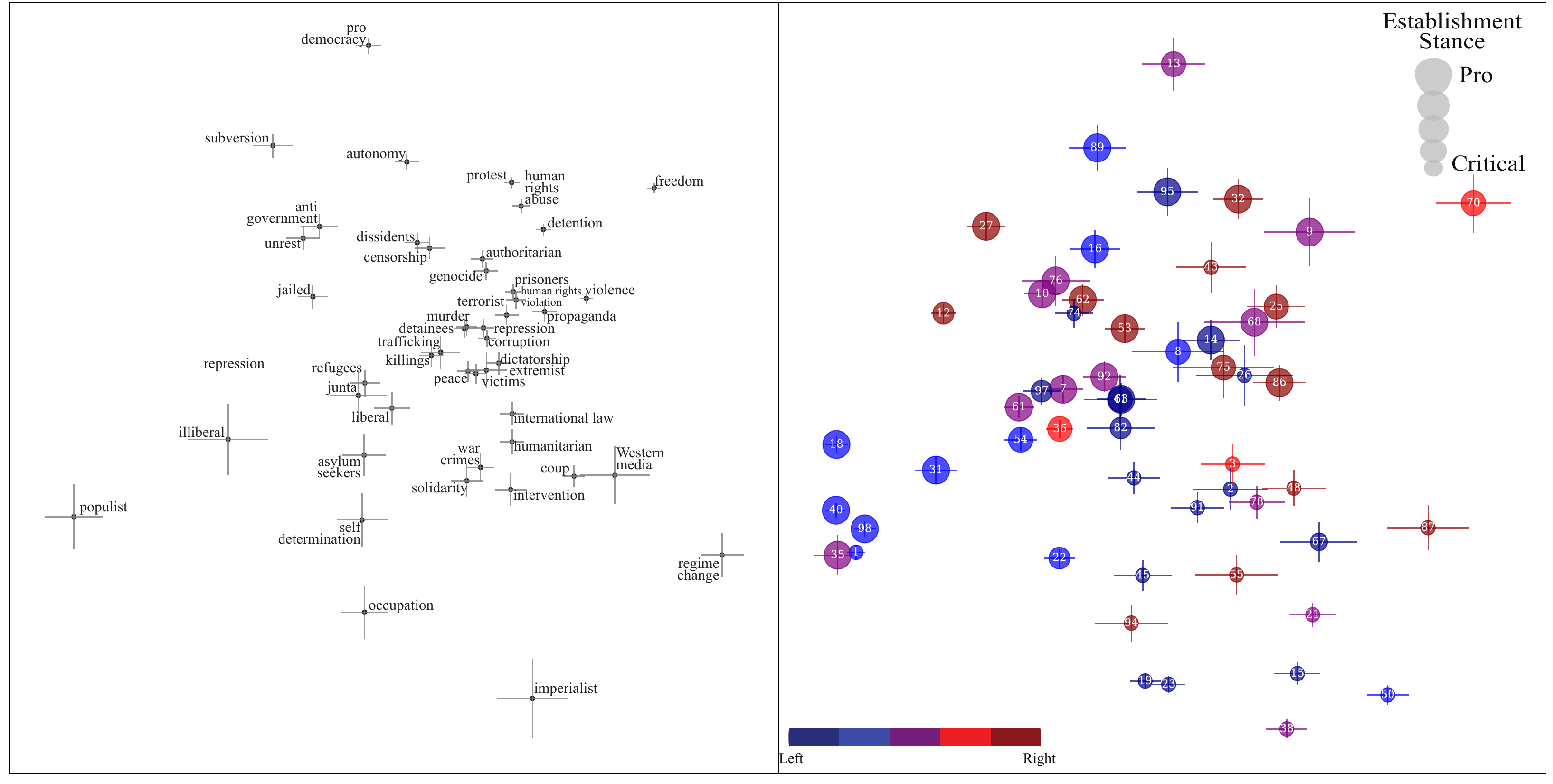}\label{humanrights}
\end{figure}

\begin{figure}[h] \hglue-4cm \caption{\bf Israel bias}\includegraphics[scale =.4]{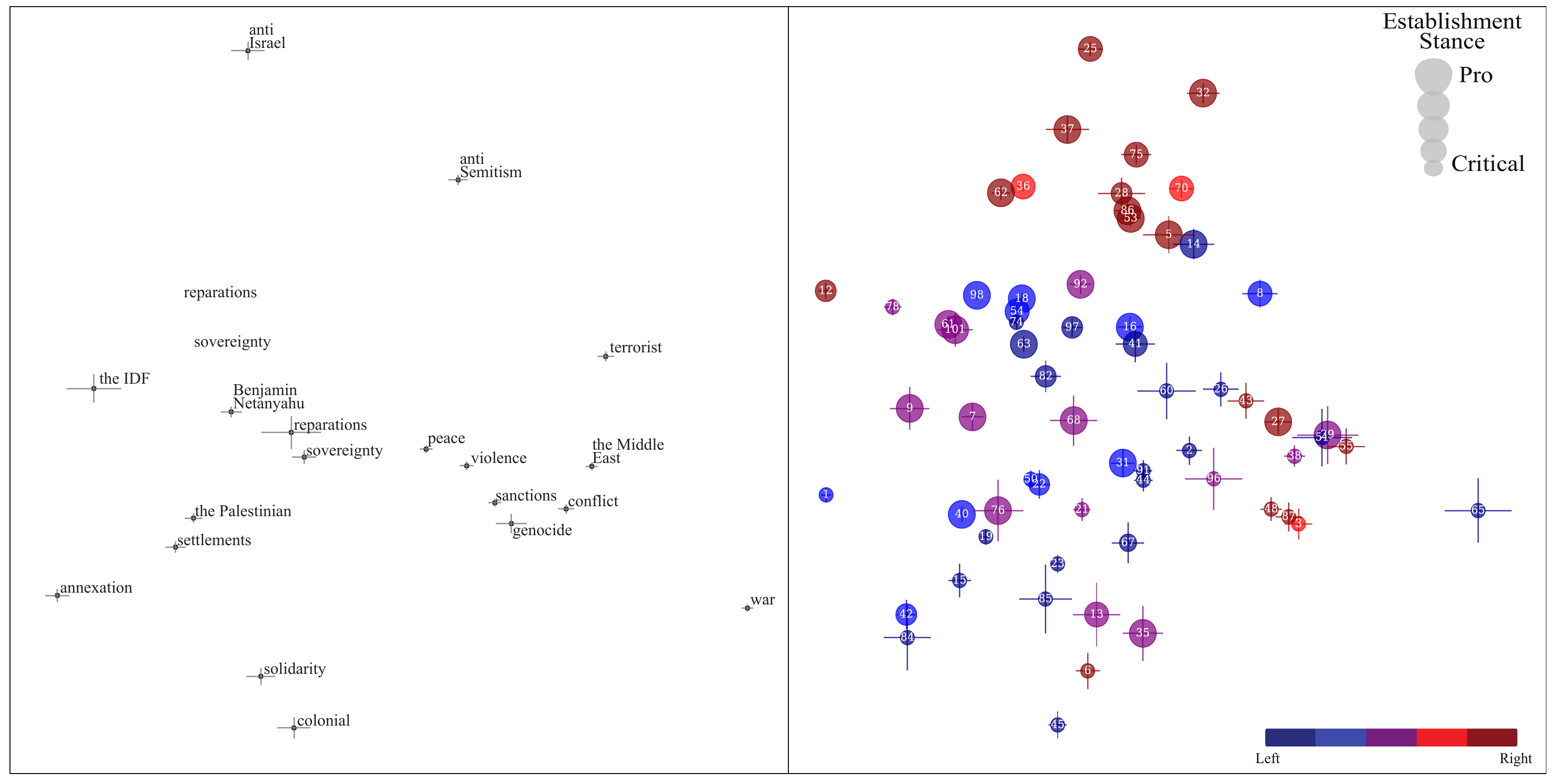}\label{ISR}
\end{figure}

\begin{figure}[h] \hglue-4cm \caption{\bf{Private Finance bias}}\includegraphics[scale =.4]{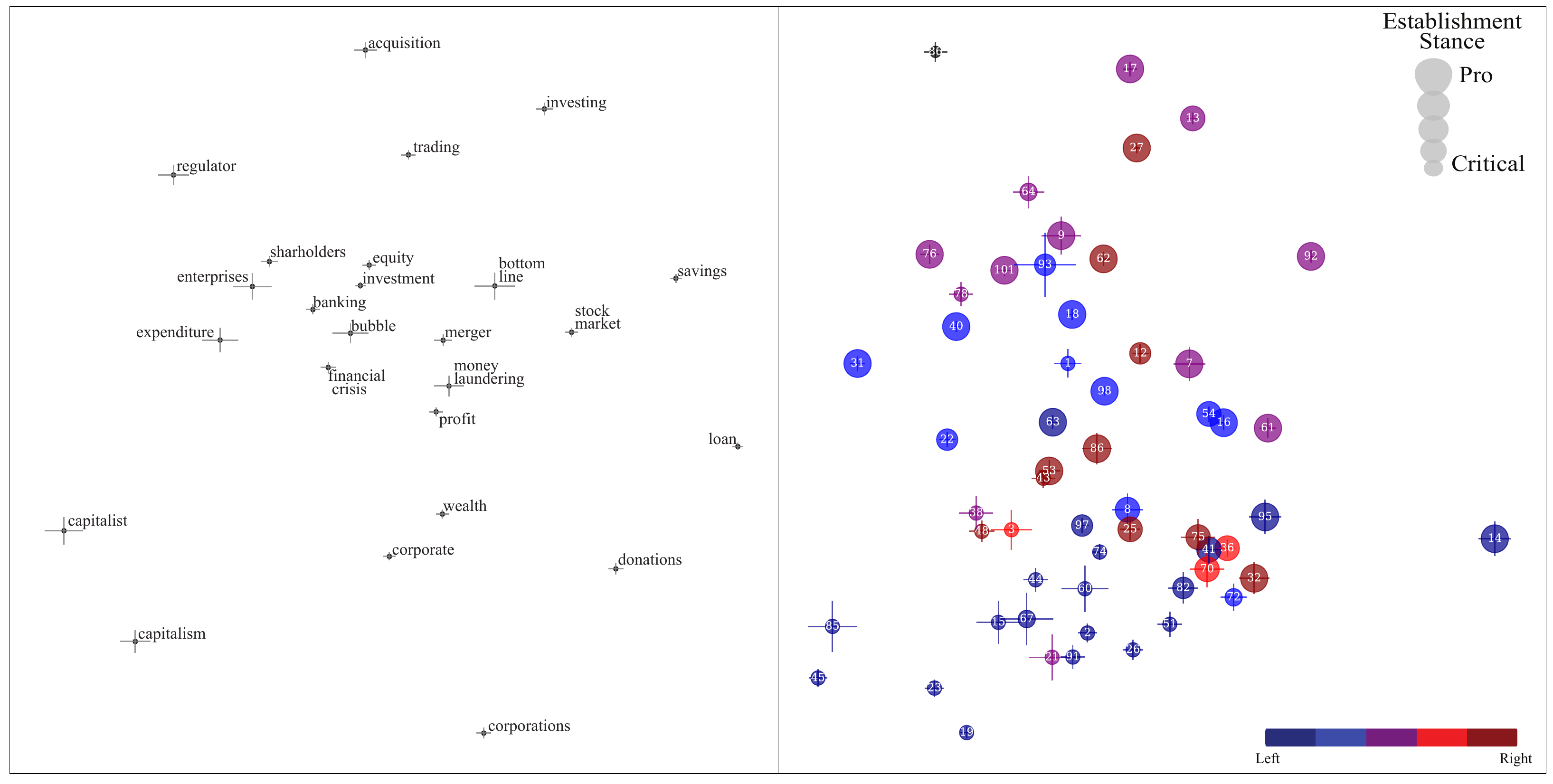}\label{private_finance}
\end{figure}

\begin{figure}[h] \hglue-4cm \caption{\bf{Native americans bias}}\includegraphics[scale =.4]{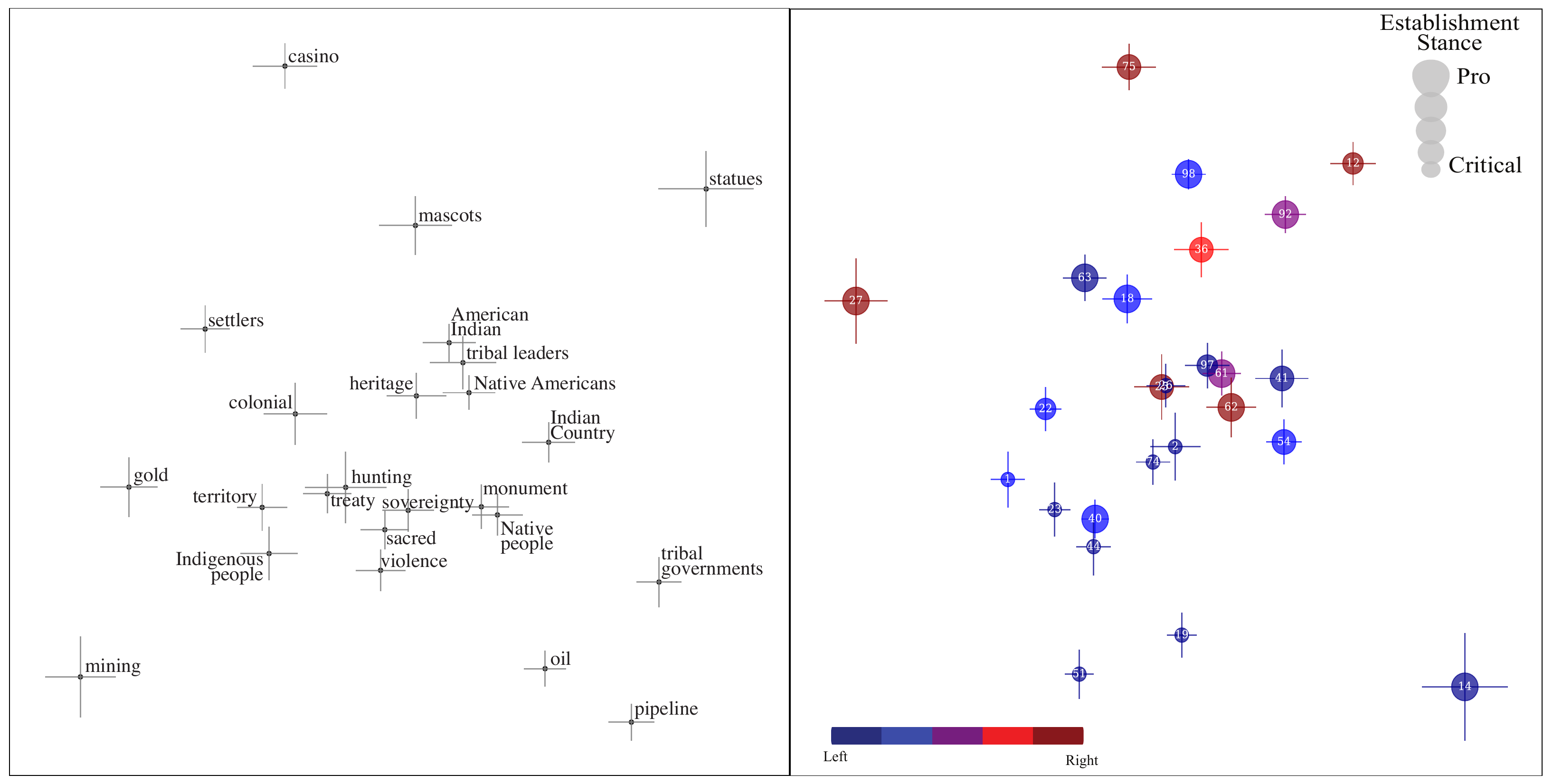}\label{native_americans}
\end{figure}

\begin{figure}[h] \hglue-4cm \caption{\bf{Oil bias}}\includegraphics[scale =.4]{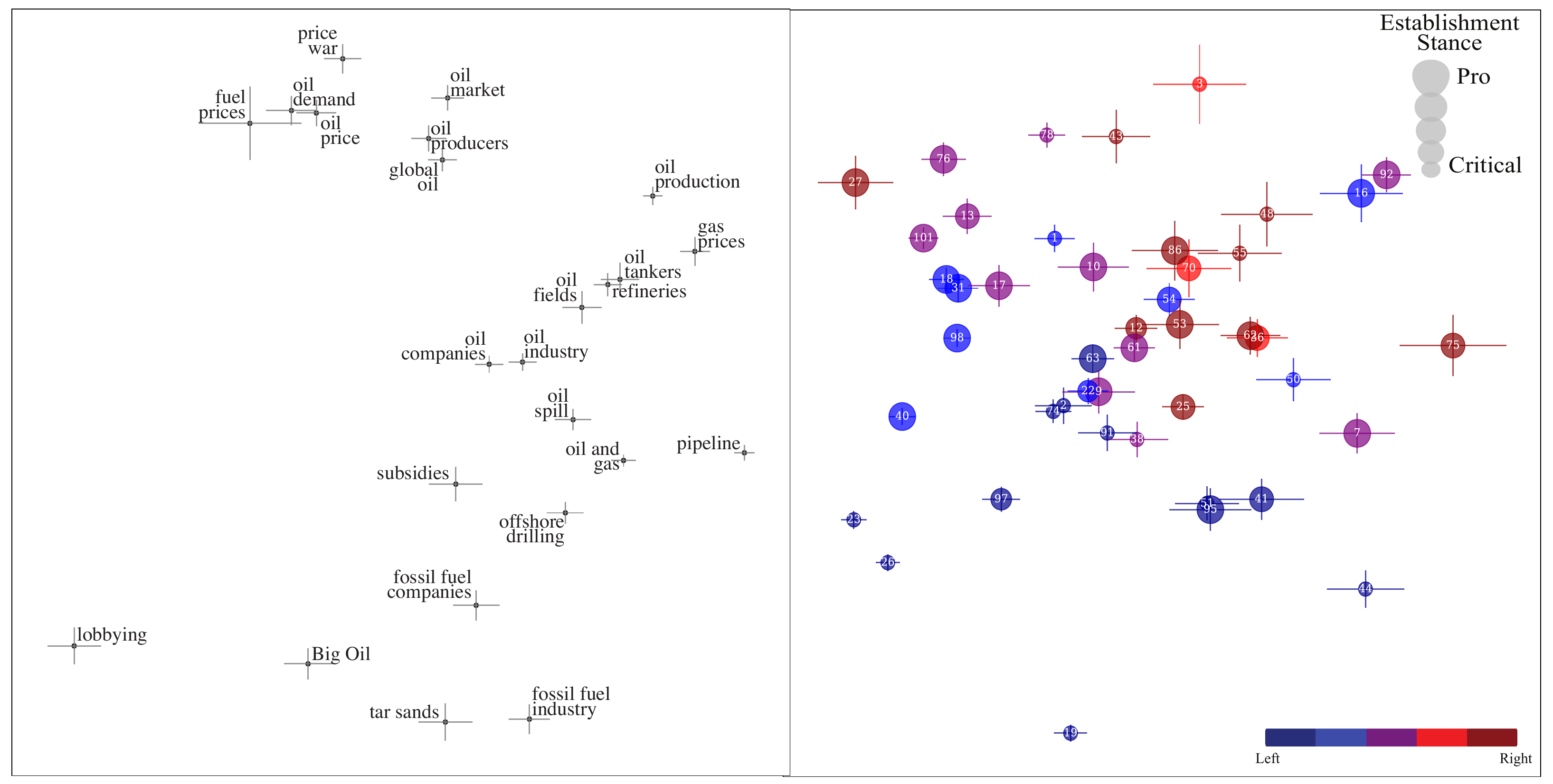}\label{oil}
\end{figure}

\begin{figure}[h] \hglue-4cm \caption{\bf{Prisons bias}}\includegraphics[scale =.4]{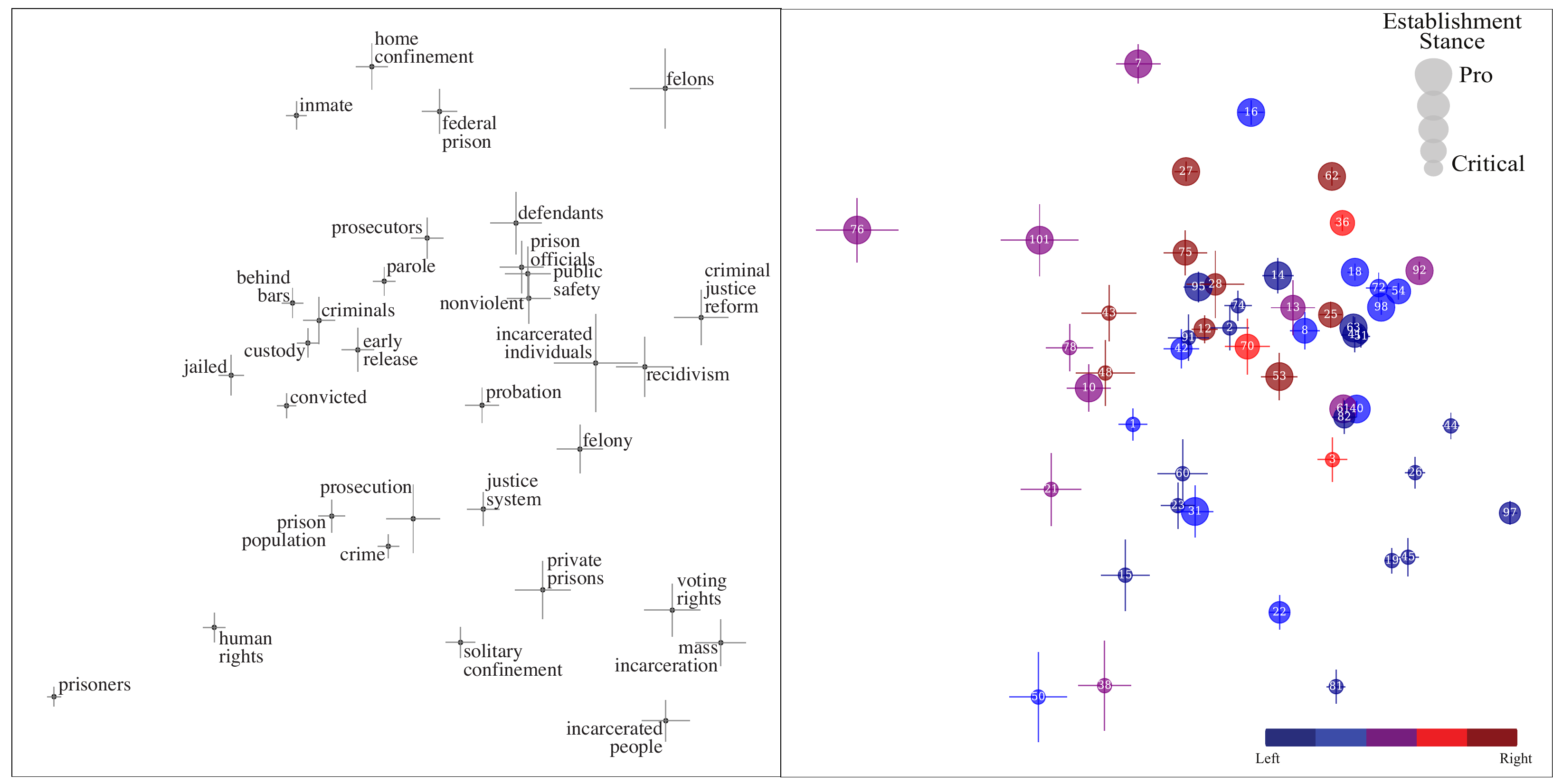}\label{prisons}
\end{figure}

\clearpage
\begin{table}[h]
\caption{\bf{Topic relevance weights for the left-right and establishment topic clusters}
\label{RelevanceTable}}
\begin{tabular}{|l|c|l|c|}
\hline
Topic Component & Left-Right Weight & Topic Component & Establishment weight\\\hline
affirmative action 1 & 0.271 & Venezuela   2        & 0.265 \\
BLM 1       & 0.245 & military spending 2  & 0.263    \\
Palestine 1          & 0.235 & Palestine 2          & 0.257               \\
guns 1               & 0.233 & Yemen 2              & 0.235               \\ 
tech censorship 1         & 0.230 & police 2             & 0.228               \\ 
Venezuela 1          & 0.226 & prisons 2            & 0.218               \\ 
abortion 1           & 0.224 & private finance 2    & 0.216               \\ 
Russia 1             & 0.202 & nuclear weapons 2    & 0.215               \\ 
Yemen 1              & 0.200 & Native Americans 2   & 0.194               \\ 
human rights 1       & 0.200 & human rights 2       & 0.188               \\ 
climate change 1     & 0.184 & China 2              & 0.187               \\ 
sexual harassment 1  & 0.181 & Russia 2             & 0.180               \\ 
US immigration 1     & 0.170 & nuclear weapons 1    & 0.178               \\ 
church state 1       & 0.166 & Israel 2             & 0.173               \\ 
police 1             & 0.163 & espionage 2          & 0.153               \\ 
church state 2       & 0.160 & Islam 2              & 0.152               \\ 
Judaism 1            & 0.160 & espionage 1          & 0.148               \\ 
Israel 1             & 0.158 & public finance 2     & 0.145               \\ 
universities 1       & 0.151 & Islam 1              & 0.136               \\ 
Judaism 2            & 0.151 & private finance 1    & 0.135               \\ 
oil 1                & 0.128 & BLM 2       & 0.133               \\ 
tech censorship 2         & 0.117 & oil 2                & 0.132               \\ 
climate change 2     & 0.117 & US immigration 2     & 0.130               \\ 
universities 2       & 0.094 & guns 2               & 0.090               \\ 
public finance 1     & 0.074 & abortion 2           & 0.088               \\ 
&& China 1              & 0.083               \\ 
&& affirmative action 2 & 0.066               \\ 
&& sexual harassment 2  & 0.054               \\ 
&& Native Americans 1   & 0.053               \\ 
&& prisons 1            & 0.049               \\ 
&& military spending 1  & 0.048               \\ \hline
\end{tabular}
\end{table}

When we performed the generalized singular value decompositions for each topic, we had the freedom to choose both the sign of each plotted component and whether we numbered it 1 or 2.
To eliminate these ambiguities and standardize the components, we the automatically flip signs such that all topics $k$ in Cluster 1 have
$E_{1k}>0$ and all topics in Cluster 2 have $E_{2k}>0$, and numbered the two components as follows.
The two components each have a relevance weight as shown in the \tabl{RelevanceTable}: 
the one with the largest relevance weight is numbered ``1" if it's a left-right component and ``2" otherwise; the second component gets the opposite number.

\end{document}